\documentstyle[astrobib,psfig,epsf]{mn-ab}

\def\rvir{R_{\rm vir}}
\def\rt{R_{\rm t}}

\def\omm{\Omega_{\rm m}}
\def\oml{\Omega_{\Lambda}}
\def\Dvir{\Delta_{\rm vir}}
\def\lcdm{$\Lambda$CDM}
\def\hmpc{\ h^{-1}{\rm Mpc}}
\def\rs{R_{\rm s}}
\def\Rs{R_{\rm s}}
\def\Msun{M_{\odot}}
\def\hMsun{h^{-1} \Msun}
\def\hmsun{h^{-1}M_{\odot}}
\def\hsolmass{h^{-1}M_{\odot}}
\def\kms{\ {\rm km\,s^{-1}}}
\newcommand{\nbody}{$N$-body~}

\title[Interaction rates of dark-matter halos and subhalos]
      {Interaction rates of dark-matter halos and subhalos}
\author[Kolatt et al.]
{T.S. Kolatt$^{1,2}$, J.S. Bullock$^{3,2}$, Y. Sigad$^1$, A.V.
Kravtsov$^{3,4}$, \cr
A.A. Klypin$^4$, J.R. Primack$^2$ \& A. Dekel$^1$ \\
$^1$Racah Institute of Physics, The Hebrew University, Jerusalem, 91094,
Israel\\
$^2$Physics Department, University of California, Santa Cruz, CA 95064
USA\\
$^3$Department of Astronomy, Ohio State University, Columbus, OH 43210 USA\\
$^4$Astronomy Department, New Mexico State University, Box 30001,
            Dept. 4500, Las Cruces, NM 88003-0001 USA}

\begin{document}

\maketitle

\begin{abstract}

The interaction rates of dark-matter halos and subhalos, including 
collisions and mergers, are computed using high-resolution cosmological
\nbody\ simulations of the $\Lambda$CDM model.
Although the number fraction of subhalos of mass $>2\times10^{11}\hsolmass$ is 
only $\sim10\%$, we find that the interaction rate of such subhalos is 
relatively high because they reside in high density environments.  
At low redshift, the subhalo collisions dominate the total collision rate,
and even at $z=3$ they are involved in more than 30\% of all collisions. About 
40\% of the ``major" mergers (those of mass ratio $>0.3$) are between subhalos.
Therefore subhalo interactions must be incorporated in models of structure
formation.  

We find that the collision rate between halos in physical density units,
is $\propto (1+z)^\delta $, with $\delta = 3-4$, in agreement 
with earlier simulations and most observational data.

We test previous analytic estimates of the interaction rates 
of subhalos based on statistical models, which could
be very inaccurate because of the small number of subhalos and the 
variation of conditions within small host halos. We find that,
while such statistical estimates may severely overestimate the rate 
within hosts $< 10^{13} \hsolmass$, typical of high redshifts, 
they are valid for larger hosts regardless of the number of 
subhalos in them. We find the \citeN{makino-hut} estimate
of the subhalo merger rate to be valid for hosts $\ga 6\times10^{11}\hsolmass$ 
at all redshifts.

The collision rate between subhalos and the central
object of their host halo is approximated relatively well using the
timescale for dynamical friction in circular orbits.
This approximation fails in $\sim 40\%$ of the cases,
partly because of deviations from the assumption of circular orbits
(especially at low redshift) and partly because of the invalidity
of the assumption that the subhalo mass is negligible (especially at
high redshift).
\end{abstract}

\section{Introduction}
\label{sec:intro}

Interactions among dark matter halos and the galaxies that reside in
them are inevitable in any cosmological scenario of hierarchical
structure formation and have important observational implications.
Halo interactions cause tidal stripping and harassment \cite{moore:98}. 
They promote the exchange of angular momentum 
(e.g., \citeNP{barnes-efstathiou}),
trigger star bursts (e.g., \citeNP{mihos:94a,mihos:94b}), 
mix different stellar populations \cite{larson-tinsley:78},
produce tidal tails \cite{toomre:72,dubinski:99,springel-white},
and change the galaxy morphology (e.g., \citeNP{oemler:92}).
The tidal stripping and harassment that occur during these
interactions enrich the intergalactic medium with
processed elements \cite{kolatt-od}, 
disrupt existing galaxies, build new ones, and 
heat up the galaxy interior matter.

Galaxy interactions are observed directly 
(e.g., \citeNP{int-groups,int-faint,int-clusters,int-mds}) and indirectly
by statistical means and through their outcomes 
\cite{abraham:97,conselice:00,kwu-phd}.
The interaction rate rises as function of redshift 
\cite{le-fevre:00,patton:00,patton:98,carlberg:1990,zepf+koo}
(but cf. \citeNP{carlberg:00})
and the density of the environment.
Since the interaction
rate is roughly proportional to the number density squared times 
the relative velocity, groups and clusters of galaxies show a higher specific
interaction rate (per object) than the specific rate in the field.

The details of the physical processes that take place during and as a
result of these interactions are not well understood.
Gas dynamics, star formation and supernova feedback hinder 
a direct understanding of how such interactions influence the
process of galaxy formation in general.  
In order to overcome this difficulty, 
semi-analytic investigations \cite{kwg:93,cafnz,sp:99,spf:00}
use simplified models for the merger rate of substructure 
(including the role of dynamical friction) 
in order to explore the possible implications of these events. 
These simplified models should be tested and improved by comparison
with realistic cosmological simulations. 

But
even before addressing the additional physics involved in galaxy interactions,
the halo gravitational interaction rate alone is too complicated to calculate 
analytically. The finite halo size and density profile, the many body 
interactions, the varying background potential all have their impact on 
the interaction rate. 

The first stage of the theoretical exploration of gravitational 
halo interactions introduces the impulsive approximation 
\cite{spitzer:58}, to estimate the effect of a close passage of two 
gravitationally bound systems. 
This approximation models the heating up of each individual
system by the transfer of gravitational potential energy
to internal kinetic energy. It also models the 
disruption by stripping and the contraction of the remaining debris. 
The impulsive approximation has been tested 
\cite{richstone:75,DLS:80} in \nbody\ simulations where the effects of close
encounters on mass exchange and profile evolution were addressed.
The main objective of these investigations was 
to test mass loss due to tidal stripping.
The conclusion was that although 
the impulsive approximation works well
even beyond ``fast" encounters, it is invalid
for encounters with small impact parameters.
Later, using two-halo \nbody\ experiments, \citeN{augilar:85} found that
the impulsive approximation works only if the minimal distance between the
passing halos is bigger than $\sim 10$ effective (de Vaucouleurs) radii. This clearly prevents the use of this approximation for 
close encounters and collisions.

The evolution of many-galaxy systems (rich clusters)
was addressed by \citeN{merritt:83} 
with the inclusion of close encounters and tidal stripping, but not merging.
Merritt investigated the question of mass segregation between subhalos and the
host cluster, and found that the implied dynamical friction time is 
longer than would be 
expected without the presence of close encounters and tidal
stripping. 

Merging of dark matter halos
was investigated on the basis of simulations 
by \citeN{makino-hut} (MH),
who parameterized equal-mass mergers as a function of
impact parameter and relative velocity for different initial density
profiles.  MH also explored the parameter space of
very close passages. 

Recently, many authors have performed detailed simulations of
individual
multi-halo systems (``clusters") 
\cite{moore:96,tormen:97,tormen+b+w:97,tormen:98,ghigna:1998,van-den-bosch:99,sensui:99,okamoto-habe:99,moore:99a}.  
These investigations focused on issues
like tidal stripping, harassment, and orbital parameters.
They led to various insights about the role interactions play
within clusters, understanding
how morphology changes due to fast encounters,
how anisotropic orbits affect the final density profile,
quantifying the dynamical equilibrium state in clusters, 
and quantifying the mass fraction of subhalos with respect to the diffuse matter. 
However, systems the size of rich and poor galaxy groups have
been neglected until now.  
The present work is the first detailed investigation of
subhalo encounters within a cosmological volume simulation.

Previously, the only relevant cosmological modeling of halo interactions
has been via analytic methods such as the Press-Schechter (PS) 
formalism \cite{ps:74} and its extended versions (EPS) 
(\citeNP{lc:93,lc:94}).  
These theoretical predictions
have been compared to \nbody\ results, 
with surprisingly good
agreement (up to a factor of $\sim 2$) 
(e.g., \citeNP{gross:suites,slkd:00}).
Unfortunately EPS theory 
only accounts for mergers between distinct, virialized halos, 
and does not consider any other form of interaction.
In particular EPS does not take into account non-merger interactions
and does not provide the joint multiplicity-mass function of the
progenitors (cf. \citeNP{sk:99}).
In addition, EPS theory neglects
the spatial distribution of the existing halos, environmental 
dependence, and, most importantly, the interactions between subhalos 
within a given encompassing virialized host halo.
All of these cases 
are extremely relevant to galaxy formation and evolution. 
For example, the results of 
MH were used by Somerville et al. (2000) to include in semi-analytic 
models the process of
subhalo mergers, which they considered to be related to the Lyman break
galaxies observed at high redshift ($z\ga3$).
In \citeN{kolatt-lbg}
we emphasize the
role these collisions play, without appealing to MH but rather as identified 
in the high-resolution simulations studied in more detail here.

One may think that the shortcomings of 
the PS/EPS formalism may be overcome by using EPS to model
isolated systems and then applying kinetic theory to
virialized systems and substructure (cf. Somerville et al. 2000).
However, this approach is questionable -- first because of the
cross-talk between isolated halos and subhalos,
and then because of the limitations of the kinetic theory.

In the kinetic theory within the framework of statistical mechanics
one predicts the collision rate among many bodies in a given system.
When we calculate collision or merger rates in a gravitating
many body system, the simple approach of kinetic theory is no longer
valid. Several important differences prevent us from applying the
conventional methods of statistical mechanics: 

\begin{enumerate}

\item The gravitating systems under consideration are not necessarily in
equilibrium. 
The velocity distribution function may be anisotropic,
may deviate from the usually assumed Maxwellian distribution
for relaxed systems, and may evolve in time.
Moreover the 
number of ``bodies" (i.e., halos) is not constant;
halos may merge or dissolve and other new halos may enter the system or just
pass through it in the course of time.

\item
The ``bodies" are not equal, but exhibit a 
large range of masses.
The mass function (and thus the cross section distribution) is
complicated and evolves in time and space.

\item The ``bodies" are not point-like or rigid
bodies, nor do they retain their identity; 
``bodies" may shrink, evaporate, grow, deform, etc. 
\label{item:par_evol}

\item Gravitating systems exhibit ``gravitational focusing" \cite{BT}, i.e.,
an increase of the cross section due to the gravitational
attraction between the interacting bodies.

\item Most of the gravitational systems under consideration are embedded
within a background potential (e.g., individual galaxy halos
 in groups or clusters). This potential cannot in general be neglected.
Moreover, the interplay between the cluster or group potential and 
its subhalos (see item \ref{item:par_evol}) evolves in time.

\item The effect of the background potential sometimes adds anisotropy to
the orbits -- for example spiraling-in of satellite halos
within a host halo due to dynamical friction \cite{BT}, or radial orbits
along filaments and to the center of deep potential wells.

\end{enumerate}

For the reasons outlined above, at present
the only way to measure the halo interaction
rate and other characteristics is to resort to high-resolution \nbody\
simulations.   
There are some minimum requirements for any simulation 
used to tackle such a problem.
One needs a big enough dynamical
range in spatial and mass resolution 
for the compilation of a statistical sample of the host/subhalo 
systems within a cosmological volume.
The mass ratio between subhalos and hosts
can range from $\sim 1$ to three or four orders
of magnitude. 
The evolution of number density of the
host halos and their subhalos is strongly dependent on the 
cosmological background and its
ability to provide a fresh supply of progenitors throughout the host
evolution. In addition, spatial resolution is of paramount importance since
substructure tends to erase within the host unless the resolution allows
its clear identification. 
Another necessary condition is the
storage and analysis of many time-steps of the simulation to determine
the temporal behavior and interaction of the substructure. 

The simulations we use in the current analysis
\cite{klypin:overcoming}
meet all of these
requirements, and in doing so
represent an important advance in simulation technology.
We have designed, tested, and used a new 
hierarchical halo finder (\citeNP{bullock_thesis,profiles}
hereafter BKS+)
specifically for the goals of this paper.

In this paper we focus on the DM component alone and
analyze the interactions between halos and 
within host halos -- those that
have at least one other halo within their virial radius.
We make no discrimination by host or subhalo
masses, though we sometimes segregate the results into clusters
and groups.
A major goal of this paper is to assess various estimates for the time
scales in virialized systems that consist of distinct constituents.

In section \ref{sec:simu} we briefly discuss the \nbody
simulations and describe our strategy to find and model the
dark matter halos. We shall then define the
hierarchy we build for halos within the simulation and the various criteria for
interactions. Section \ref{sec:stat_view} prepares the ground for the
interaction rate calculation by discussing the multiplicity function of
host halos and its sensitivity to mass resolution.
The multiplicity function plays an important role in various analytical
estimates for the collision rate.
In \S{\ref{sec:rate} we shall present the collision
rate of halos, and the progenitor mass spectrum. A comparison to various
approximations for collision timescales is carried out in 
\S\ref{sec:evaluation}, and we devote \S\ref{sec:df} to a similar evaluation of the dynamical friction timescale.
We finally summarize our findings in \S\ref{sec:conc}.

\section{Simulated halos}
\label{sec:simu}

\def\lcdm{$\Lambda$CDM}
\def\Dvir{\Delta_{\rm vir}}
\def\rt{R_{\rm t}}
\def\omm{\Omega_{\rm m}}
\def\oml{\Omega_{\Lambda}}

We used the Adaptive Refinement Tree (ART) 
code \cite{kkk:97} to simulate the evolution of
collisionless DM in the ``standard" \lcdm\ model
($\Omega_{\rm m}=1-\Omega_{\Lambda}=0.3$; $H_0=100h=70$ km s$^{-1}$
Mpc$^{-1}$; $\sigma_8=1.0$).
The simulation followed the trajectories of $256^3$ particles
within a cosmological periodic box of size $L = 60 \hmpc$
from redshift $z = 40$ to the present.
A basic $512^3$ uniform grid was used, and up to six
refinement levels were introduced in the regions of highest density,
implying
a dynamic range of $\sim 32,000$.  The formal resolution of the
simulation
is thus $f_{\rm res} \approx 2 {\rm h^{-1}kpc}$, and the mass per
DM particle is $m_{p} \approx 1 \times 10^{9} \hMsun$.
We analyze 15 saved outputs at times between $z=5$ and $z=0$.

A complementary, $L = 30 h^{-1}$Mpc simulation of eight times higher mass 
resolution and two times higher force resolution is used 
to check issues of completeness and robustness to resolution. 
We analyze $\sim 10$ time-steps of this simulation in the range 
$1.7 < z < 7$.

\subsection{Model halos}
\label{subsec:model}

The identification of halos is a key feature of the analysis;
We try to make it objective and self-consistent 
when following halo interactions and
mergers (see \S\ref{subsec:find_coll}).
Traditional halo finders utilize either friends-of-friends algorithms or
overdensities in spheres or ellipsoids to identify virialized
halos. These algorithms fail to identify substructure.
We therefore have designed a new hierarchical halo finder, based on the
bound
density maxima (BDM) algorithm \cite{klypin:overcoming}.
The details of the halo finder are described elsewhere
(\citeNP{bullock_thesis},BKS+), and we summarize below only its main
relevant features.

We impose a minimum number of 50
particles per modeled halo, unify overlapping maxima,
and iteratively find the centers of mass of spheres about the
maxima.  We compute a spherical density profile about each center
and identify the halo virial radius $R_{\rm vir}$
inside which the mean overdensity has dropped to a value $\Dvir$,
based on the spherical infall model.
For the family of flat cosmologies ($\omm+\oml=1$),
the value of $\Dvir$ can be approximated by (\citeNP{bryan:98})
$\Dvir \simeq (18\pi^2 + 82x - 39x^2)/(1+x)$,
where $x\equiv \Omega(z)-1$.
In the \lcdm\ model used in the current paper, $\Dvir$ varies from
about 200 at $z\gg 1$ to $\Dvir\simeq 340$ at $z=0$.
If an upturn occurs in the density profile inside $\rvir$,
we define there a truncation radius $\rt$.

An important step of our procedure is the fit of the density profile
out to the radius $\min(\rvir ,\,\rt)$ with a universal functional form.
We adopt the NFW profile \cite{nfw:95},
\begin{equation}
\rho_{\rm{NFW}}(r) = \frac{\rho_s}{(r/\rs)\left(1+r/\rs\right)^2},
\label{eq:nfw}
\end{equation}
with the two free parameters $\rs$ and $\rho_s$ --- a characteristic
scale radius and a characteristic density. This pair of parameters could
be equivalently replaced by other pairs, such as $\rs$ and $\rvir$.
The minimum halo mass corresponding to 50 particles is
$\sim 5\times 10^{10}\hsolmass$ (for the $60\hmpc$ simulation).
The modeling of the halos with a given functional form allows us
to assign to them characteristics such as a virial mass and radius,
to estimate sensible errors for these quantities, and to remove 
unbound particles.
Using the NFW fits, we iteratively remove unbound particles from each
modeled halo and unify every two halos that overlap in their $\rs$ and are
gravitationally bound.
Finally, we look for virialized regions within $\rs$ of big halos to
enable subhalos near the centers of big host halos, e.g., mimicking
cD galaxies in clusters.

Our halo finding is complete only for halos containing more
than $\sim 200$ particles, i.e., of mass larger than 
$\sim 2\times 10^{11}\hsolmass$ (for the $60\hmpc$ simulation)
or $\sim1.5 \times 10^{10}\hsolmass$ (for the $30\hmpc$ simulation).
The resulting halo catalog is therefore gradually incomplete
below these masses, as described in Bullock (1999) and \citeN{sigad:00}
(hereafter SKB+).  This should be kept in mind below,
when we sometimes refer also to masses somewhat below 
the full completeness limit.

\subsection{Hierarchy classification}
\label{subsec:classification}

Two halos are classified as a {\it subhalo} and a {\it host}
when the center of the smaller halo lies within $\rvir$ of the larger
one.
The one exception is the case we term ``partners", in which the two
centers lie within $\rvir$ of each other but the small halo is not
fully contained within the large one and the mass ratio is smaller
than 4/3.

\begin{figure}
\centerline{\psfig{file=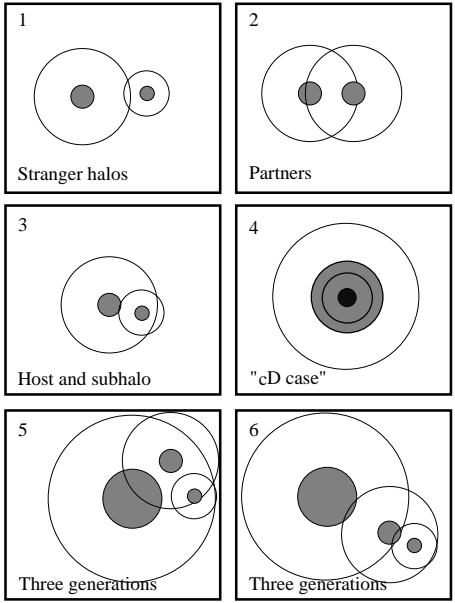
 ,height=8truecm,width=6truecm }}
\caption{Various relationships among halos and their nomenclature.
The thin lines represent $\rvir$ and the shaded regions -- $\rs$.
}
\label{fig:relations}
\end{figure}

In the current application we limit ourselves to the first level of
subhalos (but the classification scheme can straightforwardly be
extended to deal with many levels of subhalos within subhalos and with
other relations between halos).
All halos that are not subhalos are termed ``distinct" halos;
they can be either ``isolated" halos or hosts.
Halos that do not host any subhalos are termed ``simple" halos;
these can be either isolated halos or subhalos.
Figure \ref{fig:relations} illustrates the different possible
relationships and their classification.

\subsection{Interaction definitions}
\label{subsec:interaction}

The definition of halo interaction in an \nbody simulation is somewhat
arbitrary. We can easily identify violent interactions, but more subtle
phenomena are less evident. The borderline between a merger and a close
encounter, for instance, should depend upon the final bound/unbound stage of
a two body system, but finding a final bound state at some later
timestep does not necessarily
imply that a merger had occurred.
Since in reality, as in the simulations, halos can
disintegrate, or lose some of their mass during an interaction, the
decision of what we would call a merger, namely what fraction of the
progenitor halo mass should merge with another progenitor halo to form a united
halo, is a matter of definition.

In our procedure we followed these guidelines:
{(a)} All the definitions should be reproduceable within any \nbody
simulation that is being analyzed. {(b)} Avoid arbitrariness as 
much as possible in the interaction definitions. {(c)} Avoid
sensitivity of the results to particular choices of interaction
definition.
The decision to model all halos by one unique functional
form was motivated by these guidelines.

\subsection{Finding a collision}
\label{subsec:find_coll}

Interactions, by their nature, are a dynamical process. Any definition
of interaction must therefore involve the comparison of two or more
time-steps in the simulation. We hence identify and model halos in 
stored time-steps of the simulations, and compare the simulated
halos in subsequent time-steps in order to define interactions.

Our procedure for finding a collision that occurred during an interval 
$\Delta z = z_2-z_1 \, (z_2>z_1)  $ is as follows.
We list all particles that are bound to the halo $h_i$ at $z_2$. 
Particles are only listed under the highest level halo they are
bound to -- i.e., if under a subhalo, then not also its host halo.
We identify these same particles at the later time step $z_1$, 
if they are listed under any halo at this stage. 

If any of the identified
halos at $z_1$ contains particles from $h_i$ we fit an NFW
profile based on {\it these particles only}. 
The fitting parameters are listed under the halo $h_k(z_1)$.

If at the end of this process a halo $h_k(z_1)$ has more than one 
set of fit parameters from previous redshift halos, we check whether 
any pair of sets obey the ``collision criterion" (see below). 
Note that halos that at $z_2$
contained more than twice the minimal number of particles, can merge with
two separate halos at a later timestep $z_1<z_2$, 
although this is unlikely to happen.

The {\it collision criterion} is consistently defined to match the 
halo-finder;
if two density profiles, as calculated for two sets of particles,
overlap in their scale radius $\rs$, a collision
has occurred. If each of the two modeled sets is gravitationally bound to the
other modeled set then the collision is a merger. If this is not the case, the
collision is an ``unbound" collision. We use this term and
the term ``free-free collision" 
interchangeably.
Since each fit contains an error in the derived $\rs$ value, the
occurrence of a collision is up to the  combined error in $\rs$ of the
two fits.
The binding relation is defined by comparing the relative center of mass
velocity to the modeled NFW escape velocity of $h_i$ at the location 
of the center of mass of $h_j$ and vice versa.

\section{Statistical view on substructure}
\label{sec:stat_view}

Most galaxies in the local universe ($\ga50\%$) are in groups
according to various observational definitions of groups (e.g., Zabludoff \&
Mulchaey 1998 and references therein). If these definitions coincide
with physically bound virialized systems, and if one identifies the
simulated dark matter halos with galaxies (up to a certain mass) then
we should find a similar fraction of subhalos in the simulations,
provided the lower mass cutoff of the simulations accommodates the same
galaxy mass range for which the group analyses have been carried out
(see SKB+).

In our standard simulation we identify a total of $\sim 7300$ halos at
$z=0$ and $\sim 4500$ at $z=4$.
Between $5$\% to $15$\% of all
halos in the simulations we use are subhalos,
almost independent of cosmological epoch and the mass resolution
of the simulation.
According to our classification, about $10$\% of all subhalos are
subhalos of level $3$ and higher.

This leads to $\sim15-20\%$ of so-called ``galaxies" being group members
(including the host if it is not too massive) which is less than the
observed value for the simulation mass range. The
mismatch between the identified fraction of grouped galaxies versus
the same for grouped halos is resolved (SKB+) by allowing a halo tree
buildup in which one uses three times the halos' virial radius to
identify subhalos, namely replace $\rvir$ by $3\rvir$ in the definitions
(i)-(iv) in \S\ref{subsec:classification}. 
However in the current context where we are
interested in real virialized systems, we use the notion of substructure 
defined in 
\S\ref{subsec:classification}.

One of the important ingredients for the collision
probability function is the number density of subhalos within a given
halo.
This number depends first and foremost on the mass of the host 
halo, but the cosmological epoch, the power spectrum, and the host
halo shape may also affect it.
At first we focus on the host mass.
Figure \ref{fig:multiplicity_m13_z0} shows the multiplicity function
for all host halos with $M_{\rm host}>10^{13} \hsolmass$ at $z=0$.
We can identify three large clusters in the simulation ($N_{\rm sub} =
17,25,29$ at $z=0$) but we cannot detect whether the multiplicity function is
continuous throughout the subhalo number range due to insufficient
statistics. The multiplicity function is a function of the
halo/subhalo 
mass resolution. Using the same simulation 
with $> 2$ times better halo mass resolution, \citeN{colin:99}
found $4$ to $10$ times more subhalos for the same 
host halos. Note
that Colin et al. are defining halos differently and do not need to construct a
density profile for the halos in order to identify collisions. When a
mass cutoff similar to the one used here is applied to the higher mass
resolution halo catalog, a very similar multiplicity function is obtained.
For an effective mass cutoff ten times smaller than the 100\% completeness 
value used here ($2\times 10^{11} \hsolmass$) the multiplicity function
for hosts more massive than $10^{13} \hsolmass$ peaks at 4-5.
As we shall shortly see, in all of the following comparisons with model
predictions we shall use the {\it actual halos} we identify and hence we
avoid the need to use only halos that have masses
larger than our $100 \%$ completeness limit.
The completeness function for our halo finder
is presented elsewhere (SKB+).  We are $\sim 70 \%$ complete
for the smallest mass halo used in this analysis.

\begin{figure}
{\epsfxsize=2.7 in \epsfbox{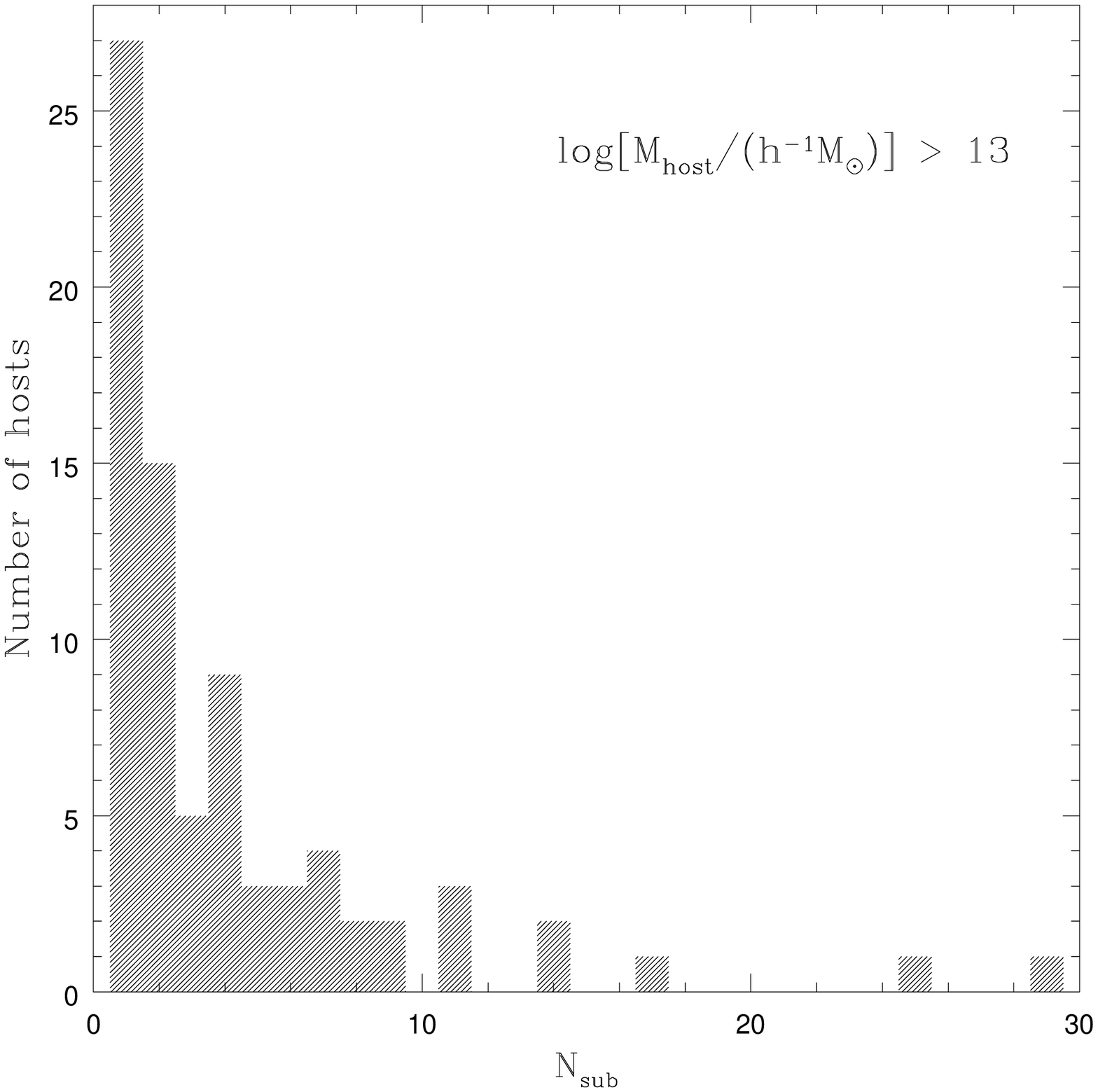}}
{\epsfxsize=2.7 in \epsfbox{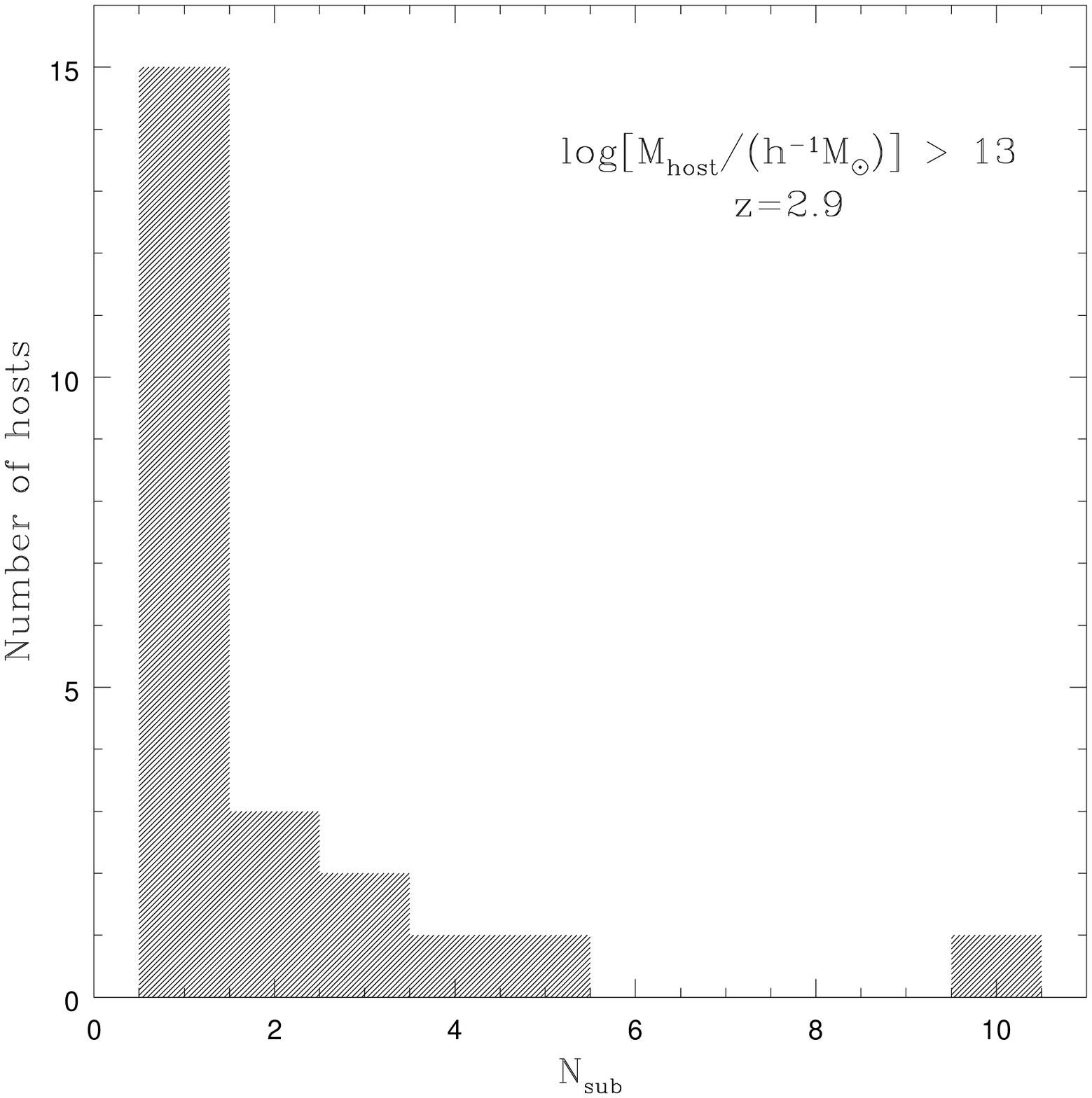}}
\caption{ The multiplicity function of all host halos with mass
$>10^{13} \hsolmass$ at $z=0$ (top) and $z=2.9$ (bottom). Note the
different axis scale. Recall that complete identification is obtained only for 
halo masses of $M>2\times10^{11}\hsolmass$.
}
\label{fig:multiplicity_m13_z0}
\end{figure}

\begin{figure}
{\epsfxsize=2.7 in \epsfbox{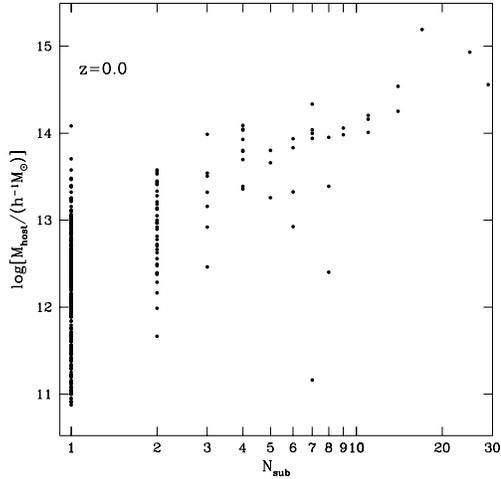}}
\caption{ Scatter plot in the multiplicity -- host mass plane for $z=0$
host halos.
}
\label{fig:multi_mass_z0}
\end{figure}

For all host halos in the current analysis, 
almost regardless of their mass, the multiplicity function
peaks sharply at the first bin, namely at a situation of a binary system
host and a single subhalo.
For comparison in Fig. \ref{fig:multiplicity_m13_z0}b 
we plot the same multiplicity function at $z=2.9$ where there are no
prominent clusters present, and the multiplicity declines sharper than
the $z=0$ function for $N_{\rm sub}=2-4$.
The multiplicity function does not seem to be a very strong function 
of the mass resolution of the simulation for high mass hosts. 
A somewhat more detailed view of the multiplicity-mass dependence is shown 
in Fig. \ref{fig:multi_mass_z0} where a scatter plot reveals the general 
expected trend of high multiplicity for more massive halos. It is clearly 
seen though that on occasion even relatively low mass halos 
($<3\times10^{13} \hsolmass$) exhibit high multiplicities.
Figures
\ref{fig:multiplicity_m12_z2.9}a and
\ref{fig:multiplicity_m12_z2.9}b show the multiplicity function of
somewhat less massive hosts at $z=2.9$ for the standard resolution and
for eight times better mass resolution, respectively.
We lowered the host mass threshold to
$10^{12}\hsolmass$ at both resolutions in order to improve on the statistics.
The limiting factor is the small number of massive halos in the small volume
of the high resolution simulation rather than the mass resolution itself.
This problem becomes more severe at higher redshifts.

\begin{figure}
{\epsfxsize=2.7 in \epsfbox{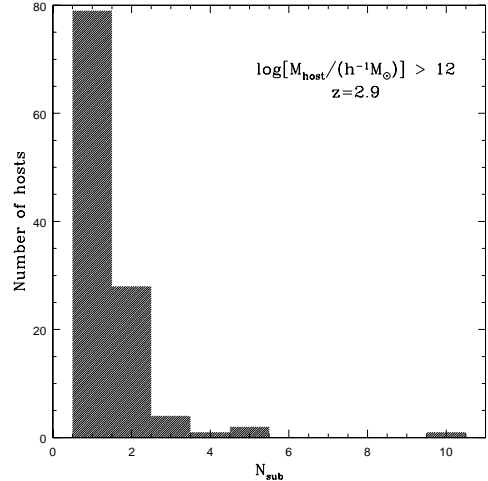}}
{\epsfxsize=2.7 in \epsfbox{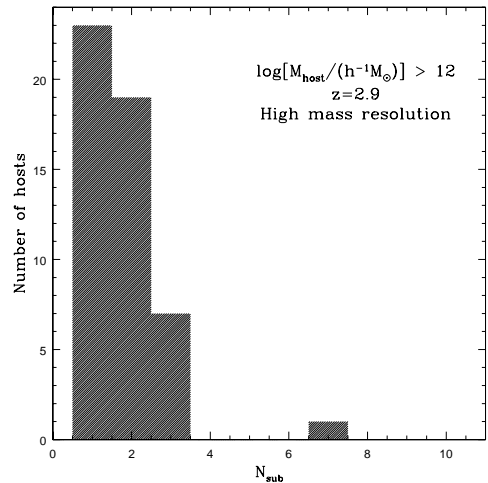}}
\caption{
The multiplicity function of all host halos with mass
$>10^{12} \hsolmass$ at $z=2.9$ for the lower resolution simulation
(top) and for the higher resolution (bottom), uncorrected for the
different volumes.}
\label{fig:multiplicity_m12_z2.9}
\end{figure}

\begin{figure}
{\epsfxsize=2.7 in \epsfbox{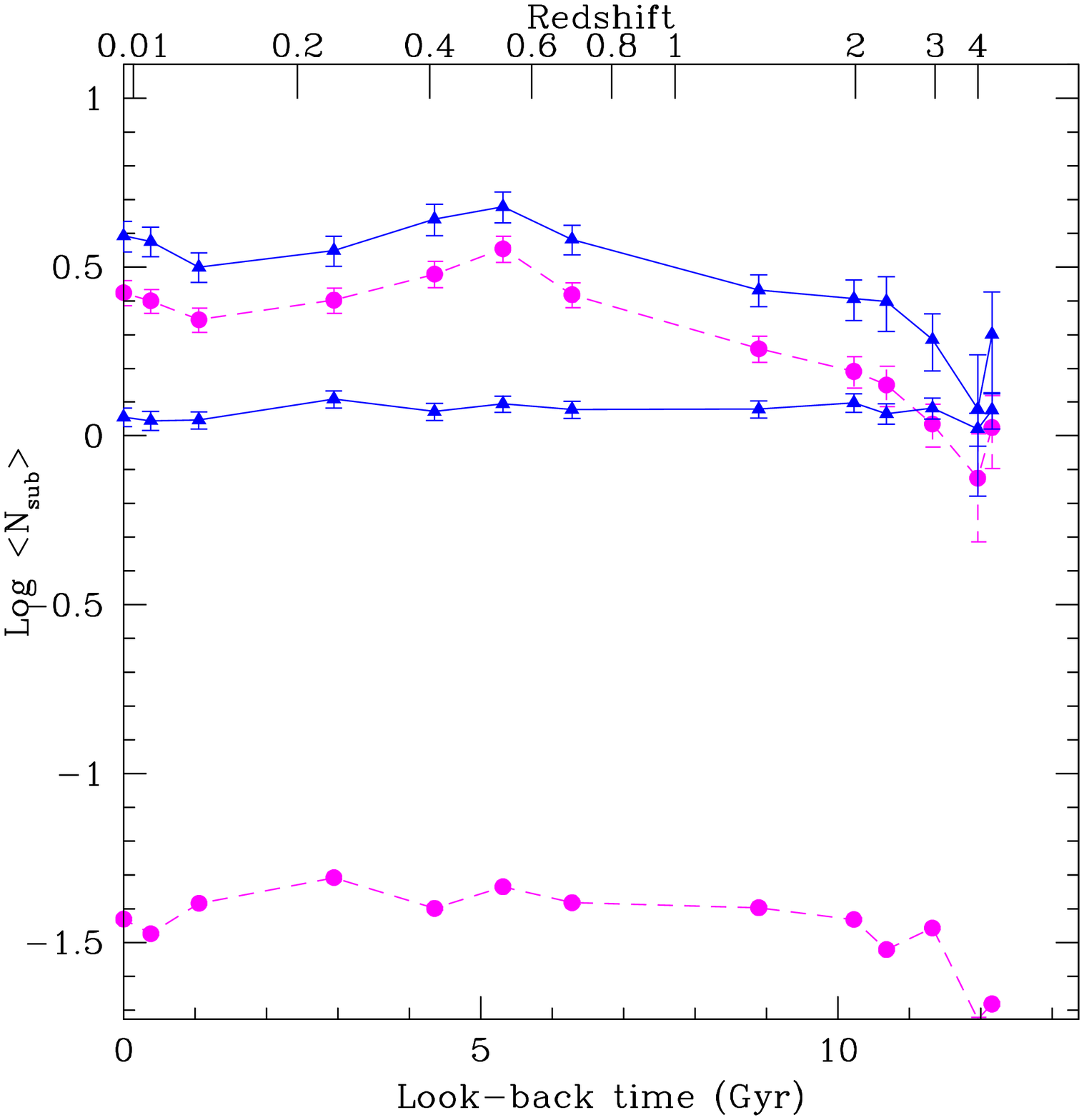}}
{\epsfxsize=2.7 in \epsfbox{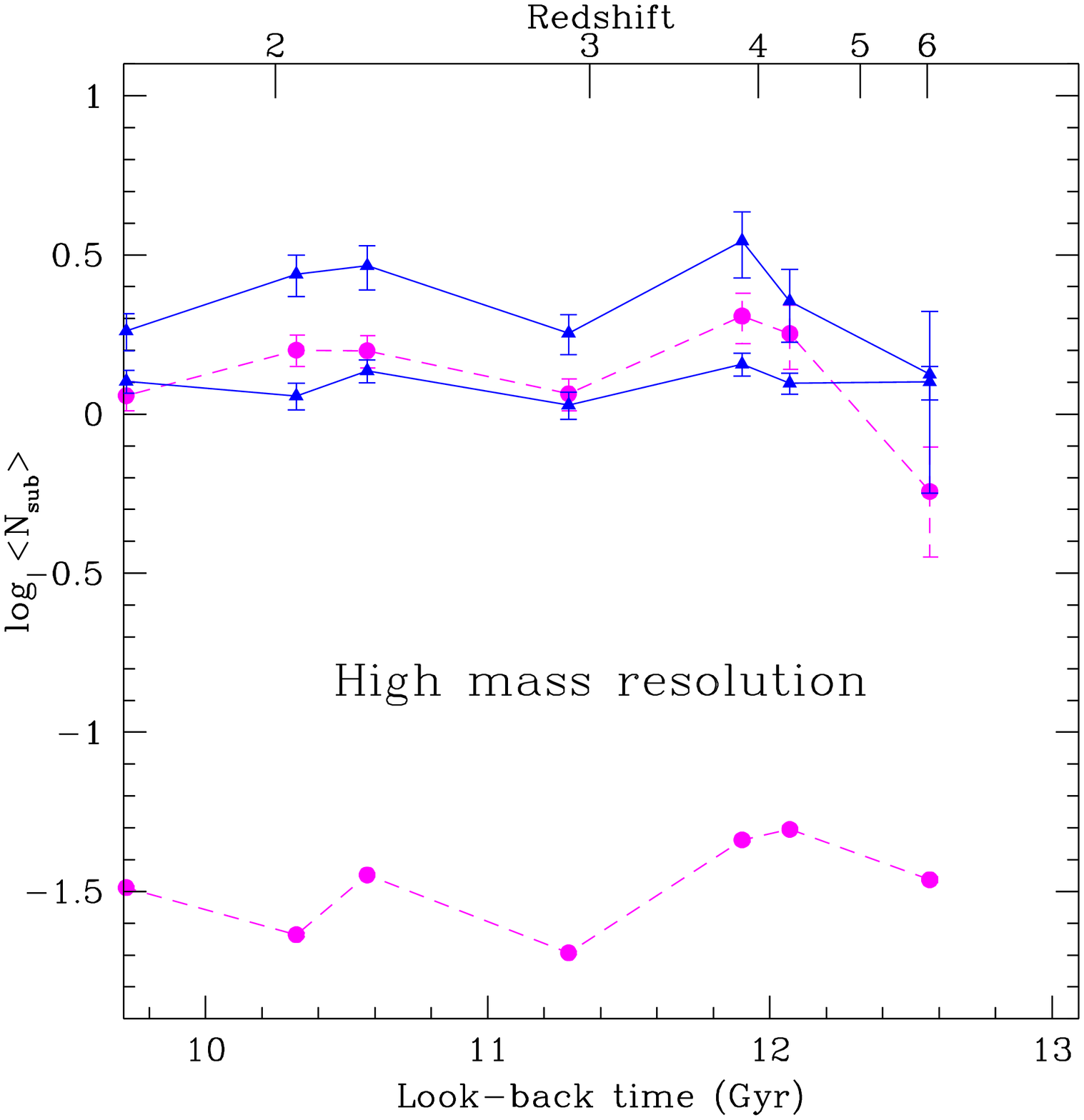}}
\caption{ The average number of subhalos per host (triangles, solid
lines) and per halo (circles, dashed lines) in two mass bins:
$\log[M/(\hsolmass)] \leq 13.0$ and  $>13.0$
(lower line and upper line respectively). Top:  standard resolution.
bottom: high mass resolution (eight times standard).
}
\label{fig:sub_per_host}
\end{figure}

Figure \ref{fig:sub_per_host} shows the average number of 
subhalos as a function of redshift for 2 mass bins,
$M$ less than or greater than $10^{13}\hsolmass$.
The dashed lines show the average number for all the halos, regardless
of whether they contain subhalos or not, whereas the solid lines focus
only on halos that contain at least one subhalo.
Most halos in the larger mass bin contain subhalos.
The average number of subhalos for host halos rises substantially
from $\sim 1.0$ to $\sim 2.0$ between $z=5$ and $z=2$ and levels to a
constant of $\sim 2.3$ subhalos per host for $z\simeq 0-1$.
In the small mass bins the common situation is one subhalo per host,
i.e., a binary system.

\section{Collision rate}
\label{sec:rate}
Recall that halos that overlap in their NFW $R_s$ parameters, and that 
mergers are bound collisions (\S\ref{subsec:find_coll}).
In addition to determining the collision rate of all halos, in
this paper we concentrate on the collision rate among subhalos, where
the multiplicity function dictates much of the characteristic behavior.
We may already guess from the lack of strong redshift evolution of the structure of 
this function
(Fig. \ref{fig:multiplicity_m13_z0}a in comparison to 
Fig. \ref{fig:multiplicity_m13_z0}b and Fig. \ref{fig:sub_per_host})
that the subhalo collision rate does not 
decrease monotonically with time at late times, 
as expected for the collision rate among distinct
halos in an expanding universe.
Figure \ref{fig:rate_sub}a shows
the collision rate as a function of look-back time or equivalently
the redshift. The overall number of collisions per Gyr per comoving volume 
is described by the solid line.  The errors are derived from the errors
in the modeled $\Rs$ of the colliding halos through the definition of 
a collision as overlapping $\Rs$ (cf. \S\ref{subsec:find_coll}).
Note that the collision and merger rates with finer time resolution were
calculated, and in every case found to be consistent with the values
that appear in Fig. \ref{fig:rate_sub} (and the following figures) within
the quoted errors. In the overlapping redshift range of the standard and
high resolution simulation, given the same mass cuts we find similar
rates within the errors. That reduces the uncertainty with regard to the
effect of the mass resolution on the calculated rate for a given mass
range (see below for a discussion of other resolution effects).
Only a fraction of these collisions occur
among subhalos or between hosts and their subhalos. The shaded area on Fig.
\ref{fig:rate_sub}a
depicts the collision rate only for collisions that involve at least one
subhalo. 
A frequently quoted statistic is the merger rate (what we call
the collision rate) in {\it physical} units as a function of redshift 
\cite{carlberg:1990,zepf+koo,patton:00,le-fevre:00}. 
Figure
\ref{fig:rate_sub_phys} shows this representation of the collision rate
where the $\propto (1+z)^{3-4}$ increase at low redshift is prominent.
Similar increases are reported by the abovementioned observational
analyses in the $z=0-1$ range, although not by Carlberg et al. (2000) --
suggesting that selection effects and the different observational tests
for collisions used by different authors may be important.

In Fig. \ref{fig:rate_sub_sub} we zoom-in onto the shaded region 
of Fig. \ref{fig:rate_sub}a. Collisions that involve subhalos can
further be divided into subclasses. The shaded region in Fig.
\ref{fig:rate_sub_sub} shows the {\it merger} (as opposed to all
collisions) rate among subhalos, while the solid thick line describes
the host -- subhalo collision rate,
namely a subhalo that got
``absorbed" by its host or went through its center without keeping its
identity. If the subhalo keeps its identity it is
classified as a ``cD" and therefore does not show up on the collision
budget. A subhalo, however, can merge with a ``cD" that resides at the
center of a host.

As expected the overall collision rate rises steeply from
$z \simeq 4$ to $z \simeq 2$, reflecting the picture of bottom up
hierarchical clustering and merger history. Below $z=1$ the rate starts
to drop gently because in a low-$\Omega_m$ $\Lambda$CDM cosmology the 
accelerating expansion halts
the collisions/mergers of isolated virialized halos. However in this
redshift range the collisions among subhalos start to become a 
substantial fraction of all collisions. By that time galaxy groups and clusters
start to form, and a significant increase in the number density of subhalos 
is obtained in these big host halos. The collision rate
increases correspondingly. The increased velocity dispersion helps the
process as well, though free-free collisions are still less probable 
than mergers. The fraction of unbound collisions
out of all collisions among subhalos remains almost constant.
Most of the erratic temporal features in the collision rate are due to
unbound collisions between a subhalo and its host.

\begin{figure}
{\epsfxsize=2.7 in \epsfbox{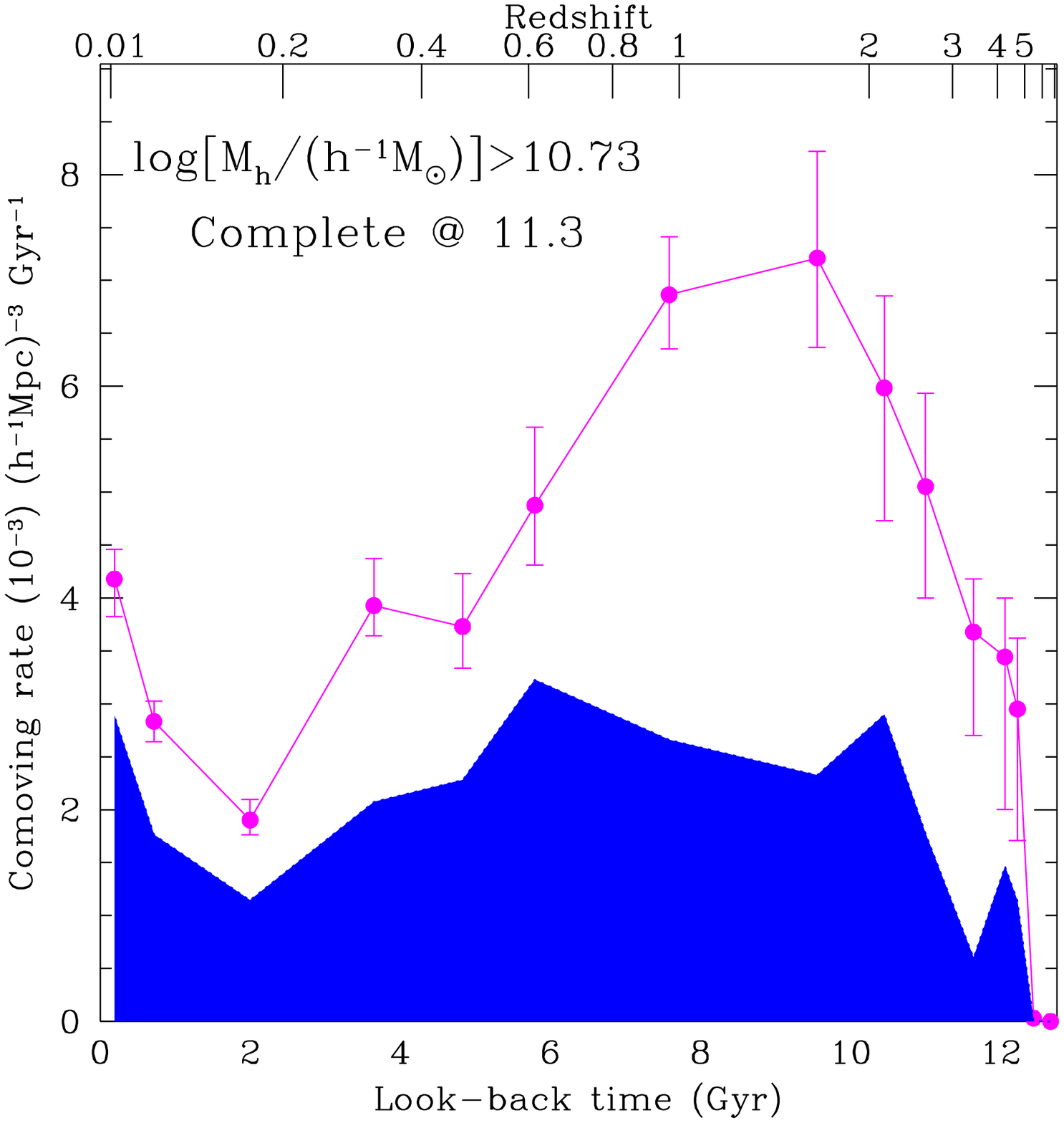}}
{\epsfxsize=2.7 in \epsfbox{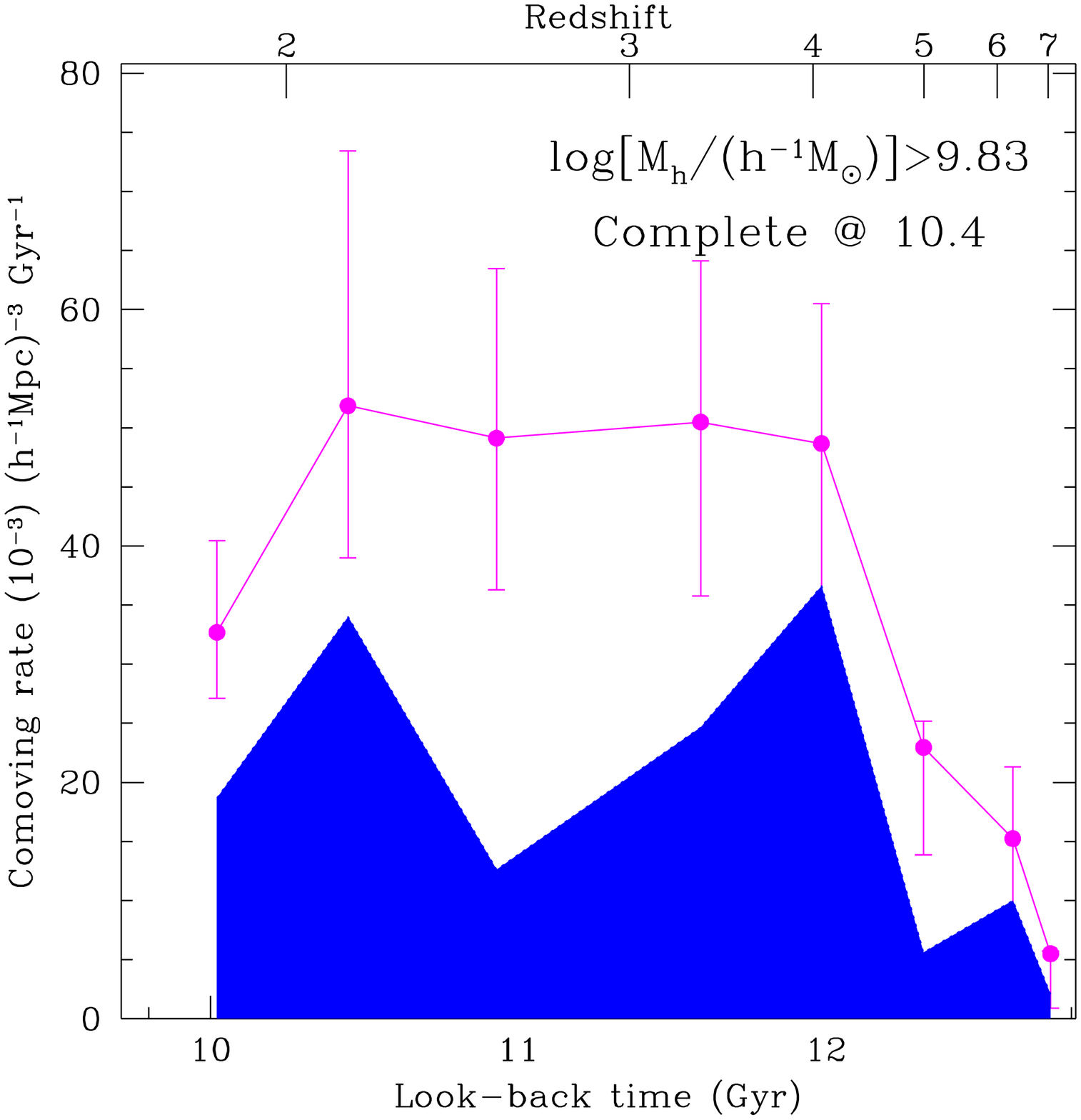}}
\caption{ The collision rate of dark matter halos with
(top) $M>5\times10^{10}\hsolmass$ (complete only above $2\times
10^{11}\hsolmass$) (bottom) $M>6\times 10^{9}\hsolmass$ (complete
only above $2.5\times 10^{10}\hsolmass$) as function of
look-back time. All collisions are described by the solid line where the
error bars are model errors (see text). The fraction of collisions that
involve at least one subhalo is marked by the shaded area.
}
\label{fig:rate_sub}
\end{figure}

\begin{figure}
{\epsfxsize=2.7 in \epsfbox{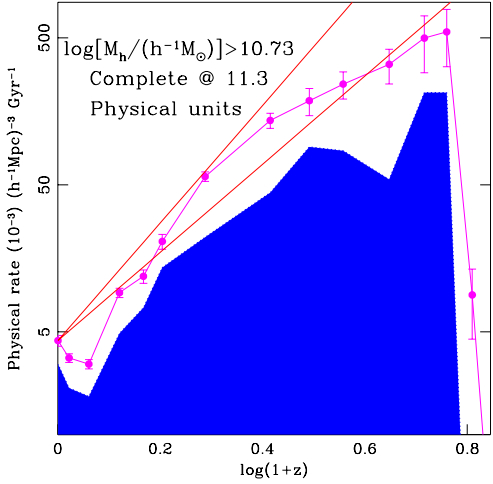}}
\caption{ Same as figure \ref{fig:rate_sub}, here in {\it physical}
units which emphasize the increase of the collision rate
and as a function of $\log(1+z)$.
The two reference lines are ${\cal R}(0)(1+z)^3$ and ${\cal R}(0)(1+z)^4$, where
${\cal R}(0)$ is the rate at $z=0$.
}
\label{fig:rate_sub_phys}
\end{figure}

\begin{figure}
\centerline{\psfig{file=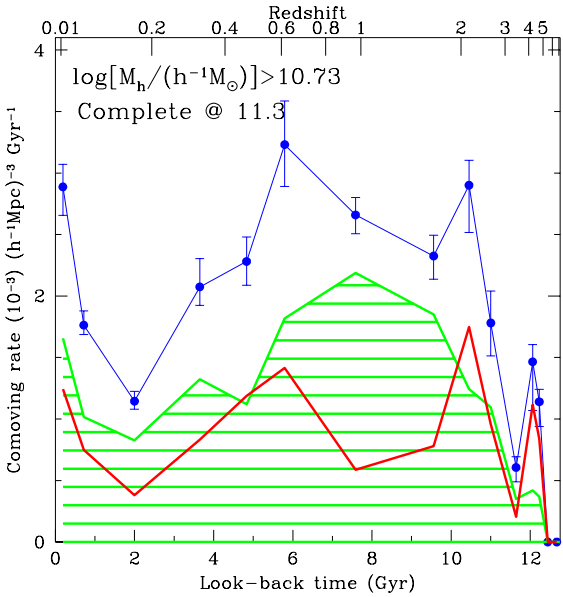
 ,height=8truecm,width=8truecm }}
\vskip0.3truecm
\caption{ The collision rate of dark matter subhalos as a function of
look-back time (thin solid line). The total number corresponds to the shaded area in
Fig. \ref{fig:rate_sub}a, where the shaded area here is the {\it merger}
rate among subhalos. The thick solid line is the collision rate between
subhalos and hosts. 
}
\label{fig:rate_sub_sub}
\end{figure}

The collision rate is sensitive to the mass resolution of the
simulation and its completeness limit. 
The higher mass resolution simulation not only
shows a much higher collision
rate (Fig. \ref{fig:rate_sub}b)
as a result of the mass function steepness, but it also shifts the
peak of the collision rate to higher redshift where small halos are
already virialized and participate in the collisions either as isolated
halos or as subhalos of hosts. Figure \ref{fig:rate_sub}b shows
the overall collision rate and subhalos collision rate at a resolution
eight times higher than the one in Fig. \ref{fig:rate_sub}a only
down to $z=1.7$. At $z=2$ the overall collision rate with
the high mass resolution is $\sim7$ times higher and the rate among
subhalos $\sim 10$ times higher. 
Recall that the product halo mass function reaches its
completeness $1$ value only at $\sim 200$ particles, i.e.,
$2\times 10^{11} \hsolmass$ for the standard resolution. This
problem is less severe than it could be because the slope of the
low mass end of the subhalo mass function (SKB+) is
only $\sim-0.7$ to $-0.5$. 
The mass spectrum of the progenitors
in the two resolutions is of course very different. 
For any specific comparisons with observations (e.g., Kolatt et al.
1999)
one has to first figure out the relevant mass range for progenitors.

\subsection{Progenitor mass spectrum}

Two important issues have to be examined with respect to the progenitor
mass distribution. It is important to learn the mass range of
progenitors in order to relate them eventually to physical entities. In
particular, whether a collision is a
``major" or ``minor" one depends on whether the progenitor
mass ratio is greater or less than some particular value, typically 0.3.

\begin{figure}
\centerline{\psfig{file=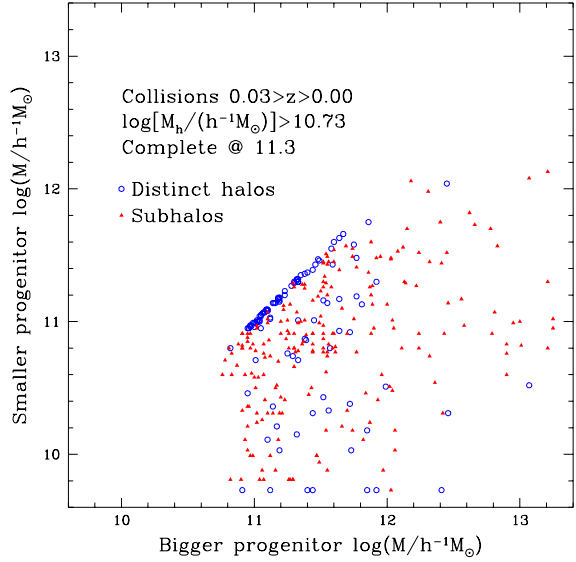
 ,height=8truecm,width=8truecm }}
\vskip0.5truecm
\caption{ Progenitor mass spectrum for collisions between $z=0.03$ and
$z=0$. Empty circles mark masses of progenitors that are both distinct
halos and solid triangles represent progenitors of collisions that involve at
least one subhalo.
Here and in the next figures, these masses are based on the particles 
that actually participate in the collision.
}
\label{fig:mdist_z=0}
\end{figure}

Figure \ref{fig:mdist_z=0} shows the distribution of masses for pairs 
of progenitors of collisions. The quoted masses of the progenitors are
calculated directly from the number of particles that took part
in the collision, which may be only fractions of the progenitor halo,
as opposed to the modeled masses (of either the progenitor halos or the
product halo). These masses can therefore be smaller than the mass
resolution limit which refers to the minimal halo mass. In most
cases these masses are a very good representation of the noisier
modeled progenitor mass, especially for large mass progenitor halos.
 
As shown in Fig. \ref{fig:rate_sub}a, the fraction of 
collisions among subhalos at low redshift is high. In Fig. \ref{fig:mdist_z=0}
we show the mass spectrum for all the collisions between $0.03>z>0.00$ divided 
into collisions between distinct halos (circles) and among subhalos (triangles).
At this low redshift almost all bigger progenitors have already 
exceeded $\sim 10^{11} 
\hsolmass$. While most of the progenitors of collisions among distinct halos
have mass
ratios of $\sim 1$, the collisions among subhalos involve a much
broader spectrum of mass ratios. The complexity of the distribution
clearly demonstrates why the numerical approach is essential. 
The mass function of subhalos cannot be derived by EPS (cf. SKB+),
nor can the progenitor mass function.
Most triangles at the right end of the figure are a subhalo -- host collision.
The pile-up of progenitors at the low mass end is a reflection of the
steepness of the mass function (cf. Somerville \& Kolatt 1999; SKB+).

Figure \ref{fig:mdist_z=2.9}a is similar to Fig. \ref{fig:mdist_z=0}
but at a higher redshift ($3.9<z<2.9$). At this higher redshift the
collisions among subhalos bear a much greater resemblance to the
progenitor mass spectrum of the distinct halos.
The higher resolution simulation (Fig. \ref{fig:mdist_z=2.9}b) at 
similar redshifts shows the general
trend of smaller mass progenitors which were not identified in the standard
resolution simulation, and similar spectrum coverage for distinct halo
collisions and collisions among subhalos. There is also a redshift
evolution (SKB+) of the subhalo mass function that folds into
the progenitor mass function at different redshift.
 
\begin{figure}
{\epsfxsize=2.7 in \epsfbox{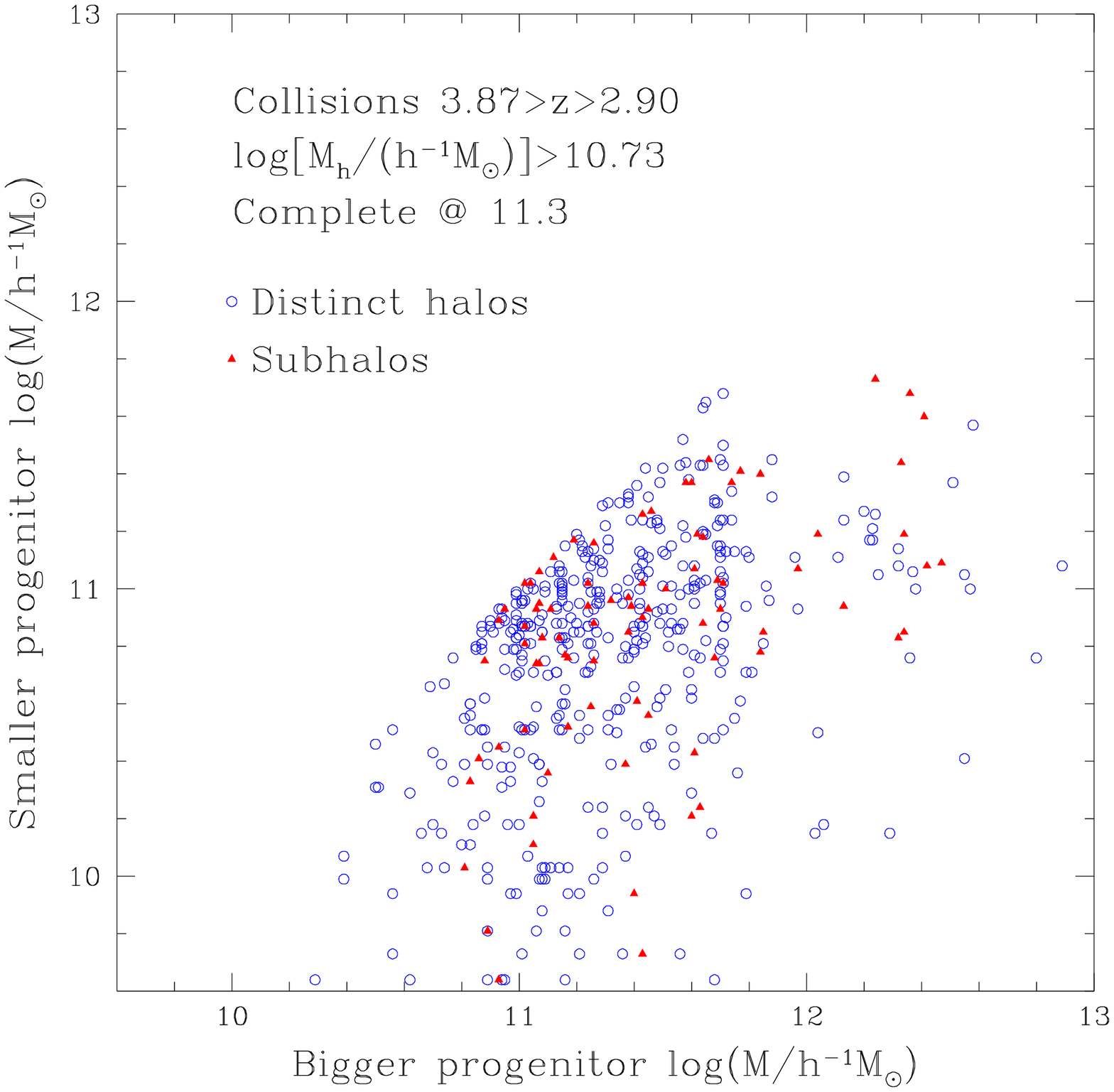}}
{\epsfxsize=2.7 in \epsfbox{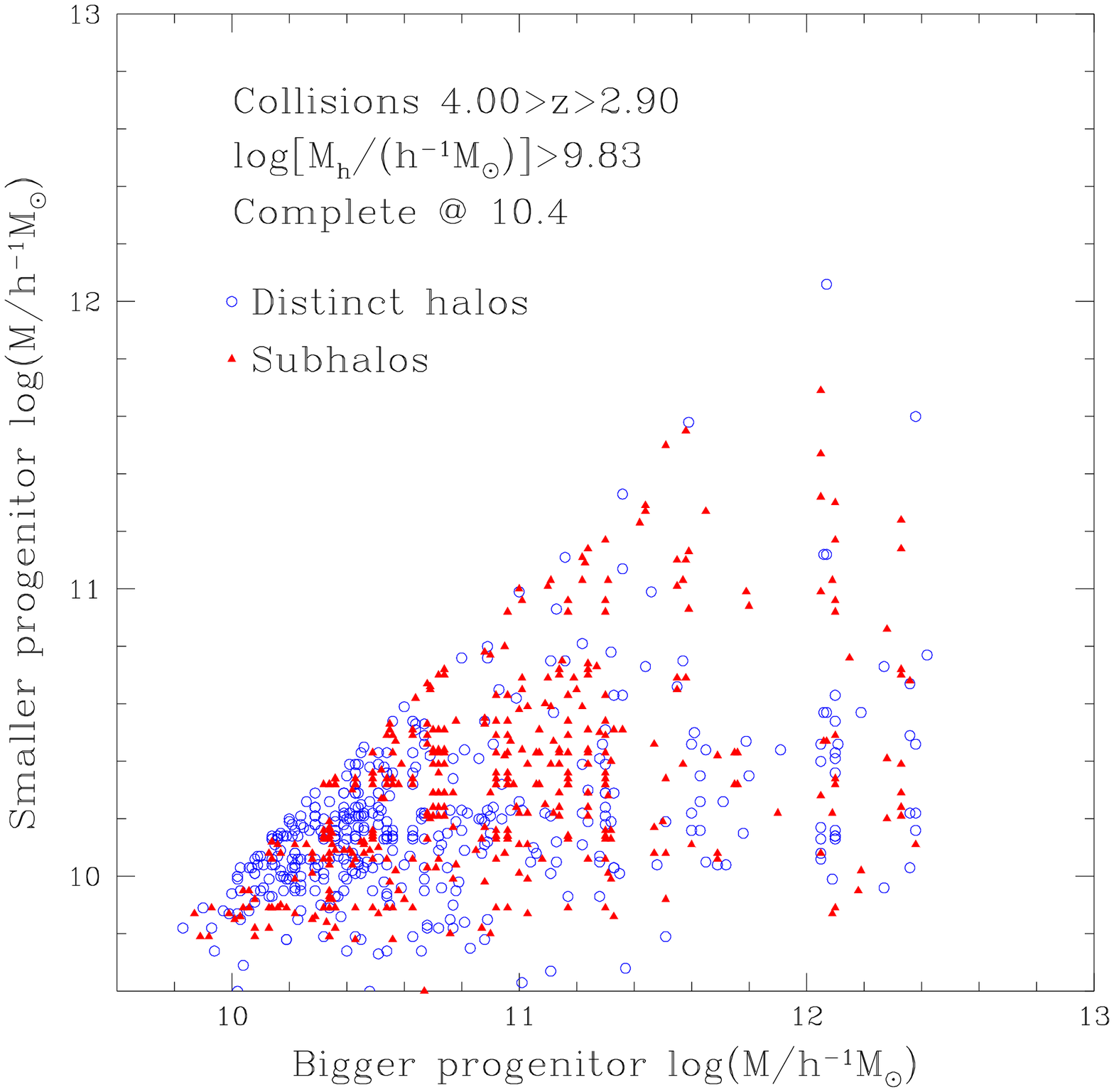}}
\caption{ Progenitor mass spectrum for collisions between $z\simeq4$ and
$z\simeq3$. Symbols are as in Fig. \ref{fig:mdist_z=0}. 
(a) The low resolution simulation (b) High resolution
simulation.}
\label{fig:mdist_z=2.9}
\end{figure}

\subsection{Major vs. minor mergers}
\label{subsec:major_minor}

The mass ratio between merger progenitors is assumed in semi-analytic
models \cite{kwg:93,cafnz,sp:99,spf:00} to determine the 
morphology of the central galaxy within the merger product.  Based on
hydrodynamic simulations \cite{barnes_rev:92}, it is assumed that 
the central galaxy resulting from a ``major merger" (dark halo mass 
ratio greater than $0.3$) 
will have a dominant spheroidal stellar component, while the 
result of a ``minor merger" of a small galaxy  (mass ratio less than $0.3$) 
with a larger disk galaxy is again a disk galaxy.
If indeed this identification is valid, we expect a higher
fraction of ``major mergers" in a high density environment, 
for consistency with the density-morphology relation \cite{dressler:80}. 
It is therefore interesting to see whether the mass ratio distribution 
alters when we turn to examine only collisions among subhalos.
 
\begin{figure}
{\epsfxsize=2.7 in \epsfbox{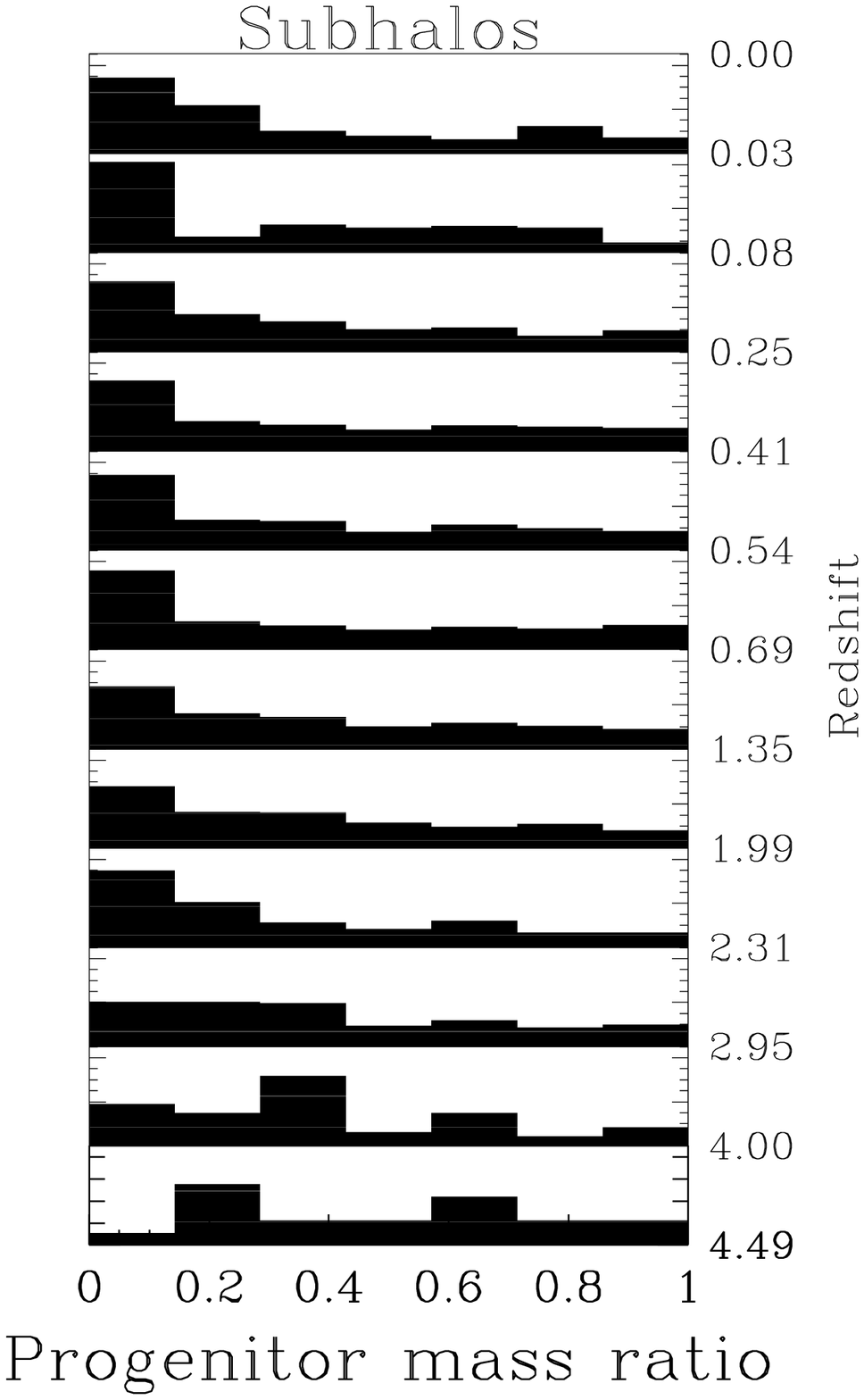}}
{\epsfxsize=2.7 in \epsfbox{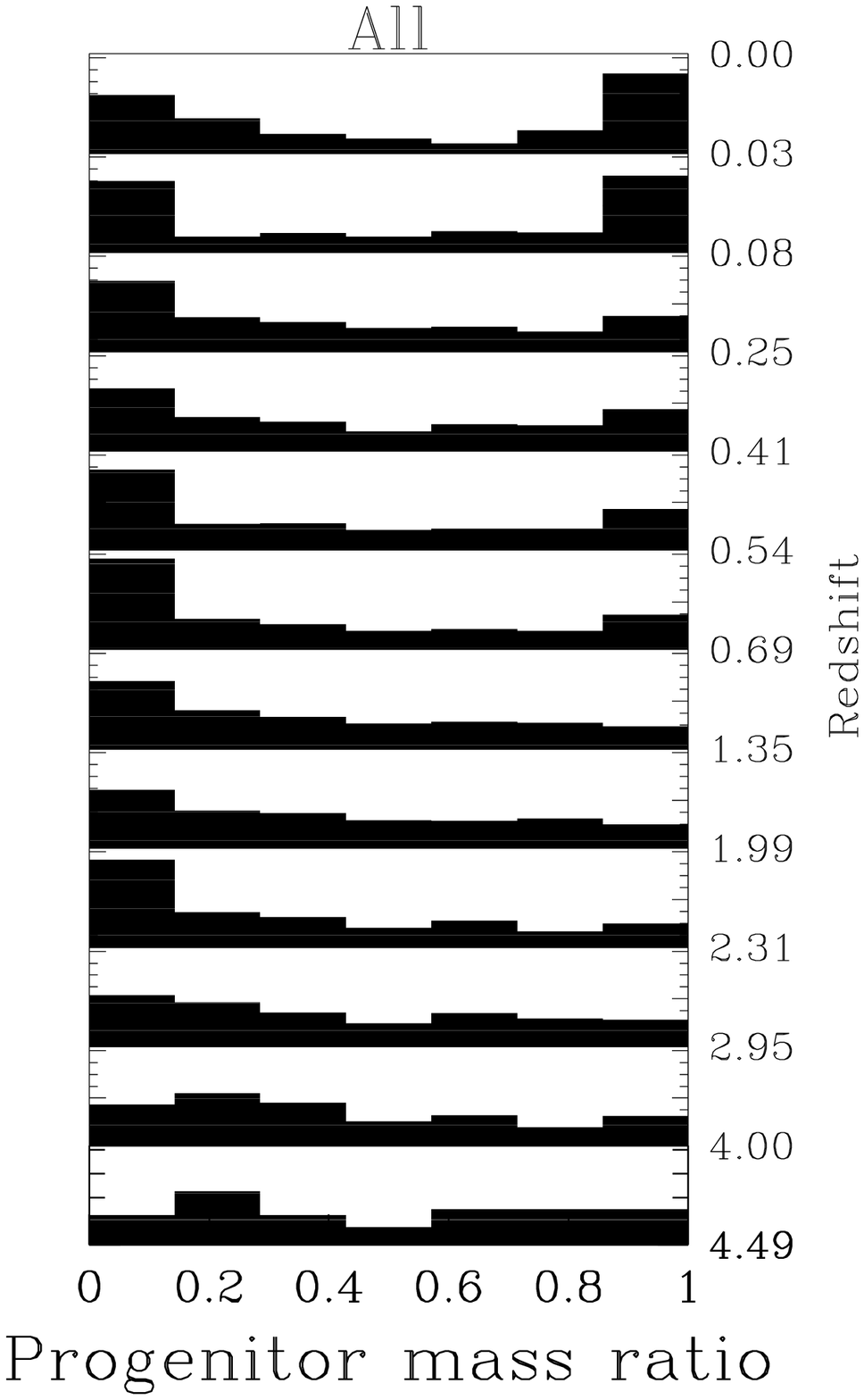}}
\caption{Probability distribution of mass ratio for progenitors of
{\it mergers} among subhalos (top) as a function of the redshift of the {
mergers}.
The redshift range is indicated on the right. The same distribution for
{\it all} mergers (bottom). The probability distributions for all
collisions are very similar.
}
\label{fig:mratio_sub}
\end{figure}

Figure \ref{fig:mratio_sub}a shows the probability distribution of the
progenitor mass ratio for mergers among subhalos, at different
epochs. At high redshift (lower part) the distribution is almost
uniform, and toward lower redshifts it tends to concentrate at low
mass ratios, namely accretion of small progenitors onto bigger subhalos.
The extended set of all collisions among subhalos shows a similar behavior. If
mergers among distinct halos are also included (Fig.
\ref{fig:mratio_sub}b) there appears at lower redshifts another
probability peak for similar mass progenitors.

If we take mass ratios $0.3$ or greater as leading to major
mergers, we find that $\sim 50-60\%$ of all mergers are major. Mergers
among subhalos start with being major in $\sim60\%$ of all cases at
$z\simeq4$ and only $\sim40\%$ by $z=0$. Similar results are obtained if
unbound collisions are included.

\section{Testing interaction rate approximations}
\label{sec:evaluation}

In section \ref{sec:intro} we have listed the various reasons for the
inadequacy of the traditional analytic calculation for the collision
rate as derived in statistical mechanics.
In this section we shall attempt to compare results from the
statistical mechanics approach and from two-halo interaction simulations
with the full \nbody results obtained here.

\subsection{Analytical calculations}

In the kinetic theory, the average number of collisions a particle
experiences per unit time is the collision rate, ${\cal R}_{coll}$. For a
system in equilibrium ${\cal R}_{coll} = n \sigma
\overline{V}$, with $n$ the number density of the
particles, $\sigma$ their collisional cross-section, and $\overline{V}$
the average relative velocity between two particles in the
ensemble. For a Maxwellian velocity distribution $\overline{V} =
v_{\rm rms} / \sqrt{\pi} $, where $v_{\rm rms}$ is the one-dimensional
velocity dispersion. If the particles interact gravitationally, and are
all identical in mass ($m$) and physical size ($r_{coll}$), 
the effective cross-section
increases (gravitational focusing, e.g., Binney \& Tremaine 1987) and
the collision rate changes correspondingly by the addition of
 $16\, G\, m\, n\, r_{coll}\,/\overline{V} $ 
with $r_{coll}^2 = \sigma/ \pi$. For a 
heterogeneous system with particles that retain their spherical
symmetry, but exhibit various $r_{coll}$ and $m$ values, we therefore 
define the average collision rate by

\begin{equation}
\langle {\cal R}_{coll} \rangle = n \left\langle \sigma \, \overline{V} + {16
\, G \, m  \, r_{coll} \over \overline{V} } \right\rangle \, .
\label{eq:rate_coll}
\end{equation}
These rates are calculated for rigid bodies with radii $r_{coll}$, and
yield a result for a {\it collision} rate as opposed to the
{\it merger} rate.

Makino \& Hut (1997, MH) derived an analytic approximation for the
merger rate, based on their binary collision simulations.
MH tested for mergers of equal-mass isolated halos of various initial
density profiles. Instead of performing a big N-body simulation, they
covered a large portion of phase space with their simulations
and derived their expression from the results.
As MH pointed out, there are numerous limitations with this approach.
The merging criterion in MH is the requirement of negative energy value
when the two systems are treated as a two body system. There is no
simple way to determine particle affiliation with halos after the
collision occurred (though the MH procedure seems very reasonable).

MH concluded that there is only a weak dependence on the initial halo
profile and that a good approximation for the merger rate in a multi-halo
system would be
\[ {\cal R}_{merger} = 
2\times10^{-3}N^2 
\left( { R_{\rm host} \over 1 \, Mpc } \right)^{-3}
\left( { R_{1/2\,{\rm sub}} \over 0.1 \, Mpc } \right)^2  \times
\]
\begin{equation}
\left( { v_{\rm rms\, sub} \over 100 \, \kms } \right)^4 
\left( { v_{\rm rms\, host} \over 300 \, \kms } \right)^{-3} 
\, {\rm Gyr}^{-1} \,.
\label{eq:mh}
\end{equation}
This approximation should be relevant for a closed system (host halo) 
of radius $R_{\rm host}$ which contains $N$ identical subhalos with 
half mass radius
$R_{1/2\,{\rm sub}}$ each. The one-dimensional velocity dispersion for the
host and the subhalo are $v_{\rm rms\, host}$ and $v_{\rm rms\, sub}$,
respectively. Note that MH neglected in this approximation the
likelihood that the central
object of their ``cluster" may have a substantially larger cross
section. Here we include it among the subhalo population of the host.
Obviously the MH approximation is only applicable to our work
if generalized to a system
with appropriate progenitor mass spectrum and varying cross sections.

\subsection{Comparison with simulations}

When we calculate the various statistical predictions, using the data of
the simulations and the formulae of the analytic or semi-analytic
approximations, we are bound to introduce an error. While the
approximations explicitly assume no evolution for the subhalo population
and neglect the host potential, the
number density and density profiles of the subhalos actually change from one
time step to another, and the background potential of the host changes
too.
In comparison to realistic \nbody results a compromise should be taken.
We would like the time lag between two compared timesteps to be as short
as possible in order to validate the no-evolution
assumption and the neglect of second order effects. On the other
hand, we would like a long enough time-step to enable accumulation of
proper statistics.
The number density of subhalos is not constant because of the entry of new 
subhalos in between the analyzed time-steps, and
subhalo destruction due to tidal stripping and galaxy harassment.

\begin{figure*}
\vskip-4truecm
\centerline{\psfig{file=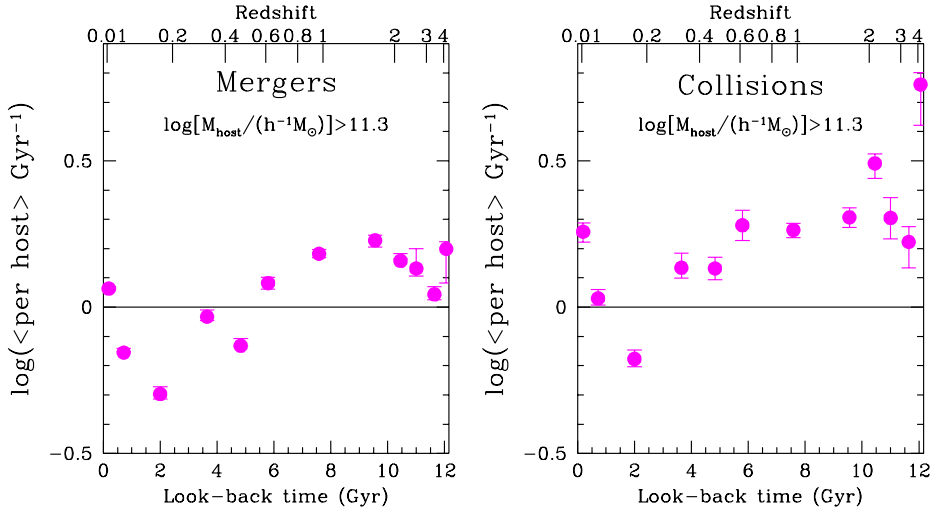
 ,height=13truecm,width=13truecm }}
\caption{The average number of (a) mergers and (b) collisions per host halo
per Gyr as a function of look-back time. The errors for collisions are
bigger due to larger analysis errors on the modeled $\rs$ of the
free-free collisions.
}
\label{fig:stat_t_blow}
\end{figure*}

We here check how well these approximations compare with the
results obtained from the simulations.  Namely, we compare
the modified kinetic theory approximation with our
measured collision rate, and the MH approximation with the
measured merger rate.
At each time step we record for each host halo the number $N$ of subhalos it 
contains,  and calculate the number density $n=N/V_{\rm vir}$ using the modeled
virial volume of the host. For consistency with our collision definition
(cf. \S\ref{subsec:find_coll}),
we take the cross section for each subhalo to be a function of
its modeled $R_s$.
The instantaneous rate for this host halo is evaluated via

\begin{equation}
{\cal R}_{coll} = 
{1 \over \sum_i W^i} \sum_i \left[ \, \pi \, (R_{\rm s}^i)^2 \overline{V} 
+ {16\, G\, M_{\rm vir}^i\, R_{\rm s}^i \over \overline{V} } \right] W^i \, ,
\label{eq:rate_host}
\end{equation}
where the sum is over all subhalos.
$W^i = R_{\rm s}^i / \sigma_{\rm R_{\rm s}}$
are the weights for the modeled $\Rs$'s of the subhalo population
($(W^i)^2$ for the simple $n\sigma \overline{V}$ estimate).
The predicted number of collisions for this host during the proper
time $\Delta t$ between subsequent registered redshifts
is obtained by
\begin{equation}
N_{coll} = N_s\, {\cal R}_{coll} \, \Delta t \, ,
\label{eq:n_coll}
\end{equation}
with $N_s$ the number of subhalos of the host halo.

For statistical reasons it is more robust to compare the global rate
rather than the rate for each individual host halo. Moreover, halos do
not keep their identity from one time-step to another. 

Figure \ref{fig:stat_t_blow} shows the average number of mergers and 
collisions 
per host halo per Gyr as a function of look-back time. The
temporal evolution of the collision rate per host shows a similar
behavior to that of the overall rate per comoving volume element 
(Fig. \ref{fig:rate_sub_sub})
with an increasing rate until $z\simeq2$ ($z\simeq1$ for
mergers) a decline until $z\simeq0.2$ and another rise thereafter.
Note that the comoving merger rate per host halo remains close to constant
throughout the redshift evolution and only varies in the range 
$0.5-2$Gyr$^{-1}$. 
The same mild variation applies to the collision rate per host halo, which
remains in the range $1-2$Gyr$^{-1}$.

\begin{figure*}
\vskip-5truecm
\centerline{\psfig{file=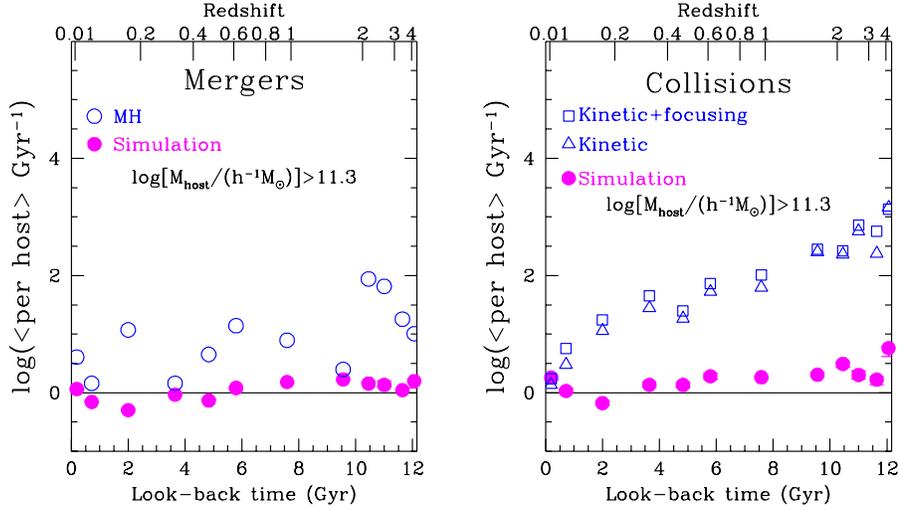
 ,height=13truecm,width=13truecm }}
\caption{
The average number of (a) mergers and (b) collisions per host halo
as a function of look-back time and redshift, in comparison with analytic
models. Solid symbols describe results from the current analysis (errors
smaller than dot size).
Empty circles on the left describe the MH predictions of merger rates
within host halos, empty symbols on the right are obtained
from the collision rate prediction of kinetic theory (triangles) and the
additional gravitational focusing (squares). 
}
\label{fig:stat_t}
\end{figure*}

Figure \ref{fig:stat_t} shows the same data as appear in Fig.
\ref{fig:stat_t_blow} in comparison to the approximation
predictions.
The empty symbols of Fig. \ref{fig:stat_t} represent the approximations
for the collision (merger) rate per host according to Eqs. 
\ref{eq:mh}-\ref{eq:n_coll}. Note that the input data are taken from the
true parameters of the simulation halos (Eq. \ref{eq:rate_host}).

Similarly, taking into account the fact that we are dealing with a realistic
mass spectrum, and not
equal-mass halos, and conforming to the half mass radius for the NFW
modeled profiles, we use the same procedure
(Eqs. \ref{eq:rate_host}, \ref{eq:n_coll}) to calculate the expected
rate \'a la MH for each host halo in the sample.
We summed all
rates (multiplied by the number of subhalos for each host) and
obtained the MH prediction for the merger rate per host halo.

It is evident from inspection of Fig. \ref{fig:stat_t} that the
theoretical approximations do not agree with the measured quantities
at high redshift, and that the disagreement tends to decrease at lower
redshifts.  Part of the reason for the
discrepancy can be read directly from Fig. \ref{fig:multiplicity_m13_z0}
--- even massive halos tend to have only a handful of subhalos
sufficiently massive to survive our cuts, and do
not reach the required large number density for statistical considerations. 
Other reasons for the discrepancy will be discussed shortly.
Note that gravitational focusing only slightly changes (by $0-50\%$) the
straightforward predictions of kinetic theory.

A way to cope with the limited number of subhalos per host is to divide 
the host population by
the host mass or the number of subhalos they host.
Although the approximations break down, they may still work for a subset
of halos where the approximation assumptions are valid. In the next section we
explore this possibility.

\subsection{Interactions as a function of $N_{\rm sub}$ and host mass}

We first test the conjecture that the governing factor for the mismatch
between the predictions of the approximations is the small number of
subhalos. 
At each given time-step the available multiplicity function does not 
allow, in general, a large enough number of host halos with high 
multiplicity. In order to
check the conjecture, though, one needs proper statistics, i.e., many host
halos of high multiplicity. A way around this difficulty is found by stacking
redshift steps together according to the multiplicity of the host halo under
consideration. 
That allows us to accumulate statistics from all time-steps
and bin the data by $N_{\rm sub}$ of the halos instead of their
collision redshift. The error we introduce by this statistics accumulation
is mainly due to different conditional mass function of subhalos at
different redshifts (SKB+).

Figure \ref{fig:stat_sub_a4} shows the average number of mergers
(collisions) per host halo per Gyr as a function of the number of their
subhalos. 
We compare only host halos with $M>2\times 10^{11}$ since at high redshift
both the host halo and the subhalo of lower masses have very low concentration 
(cf. Bullock et al. 2000), and thus may obey the ``cD" case criterion and
yield unphysically high rates.
Evidently the rate is still overestimated with all three approximations by
about one to two orders of magnitude. 
A fixed logarithmic shift of the kinetic theory model by $\sim -1.6 -
1.7$ for $N_{\rm sub} \la 25$
provides reasonably good agreement between the measured collision rate
and the calculated one, for the high resolution simulation as well.
We were able to identify the cause for the increasing discrepancy at the high
multiplicity bin. It comes from low mass hosts with high multiplicity
at $z_2$ (cf. Fig. \ref{fig:multi_mass_z0}) and some higher mass hosts 
with high multiplicity at $z_2$. For the 
high mass hosts that show the most discrepant result (a small fraction), 
the multiplicity changes dramatically from one time-step to the next, 
generally toward
lower multiplicity at the later time. The approximations 
(\ref{eq:rate_coll}) and 
(5) take $N_s$, the number of subhalos, and $n$, their number density, 
to be constant. But in these discrepant cases $N_s$ {\it decreases} due to 
tidal destruction, causing the subhalos to drop below our resolution
limit, and thus $n$ decreases as well. 

In conjunction with the additional increase in 
$\rvir$, the approximations show much higher interaction 
rates than are detected in the simulations.
The breakdown of the approximation assumptions
suggests that there might yet be a different 
factor which governs the
collision rate more strongly and causes the deviation from the
predictions.

The other governing parameter for the approximation success may be the 
host mass. 
Massive host halos on average have higher multiplicity and thus better
fulfill the statistical assumptions.
In addition to that, massive hosts are more stable and their
gravitational potential prevent subhalos from escaping the virial radius.
The competing effect is the stronger tidal force they exhibit; however, this 
is somewhat relaxed due to their typical lower concentration parameter 
(cf. BKS+) 
and hence moderate density gradients.
We use the same stacking approach by which we examined the approximation 
dependence on the multiplicity
to circumvent the difficulty of not having enough 
statistics at each time step. Here we consider mass bins, accumulating all
similar host masses from all time steps. Since the halos were chosen
under the proviso they had already virialized, similar host masses
correspond to similar velocity dispersion within them, with no regard to
the cosmological epoch, although this is not strictly valid with respect to the
number of subhalos at different epochs and their
scale radii (BKS+). 

Figure \ref{fig:stat_mass} shows the merger rate as a function of
the host mass. The simulation statistics only allow $2-3$ significant 
mass bins.
Clearly for the high mass bin the MH approximation works
well, while it overestimates the rate by a factor $\sim5$
for the low mass bin, for reasons discussed in the
previous subsection. 
The collision rate prediction from the kinetic theory works well
for the large mass bin as well and exhibits a discrepancy of
about an order of magnitude for the low mass bin. 
These results explain the large deviation from the models seen, e.g., in
Fig. \ref{fig:stat_t}. The mass function is always dominated by the low
mass end of the hosts. Naturally, this end tends to consist of less 
massive hosts
at higher redshift. Most hosts therefore do not fall in the regime where
good agreement with the models is obtained, and the average rate per host
is dictated by these.
The conclusion is that the difference between the kinetic theory
predictions with or without taking gravitational focusing into account
is not big, and they both fit the results from the simulations rather
well for masses larger than $\sim 10^{13}\hsolmass$. In the high resolution
simulation a similar trend is detected and successful prediction of all
three models can be extended due to better statistics to $M_{\rm host} 
\ga 10^{12.5}\hsolmass$.

\begin{figure}
{\epsfxsize=2.7 in \epsfbox{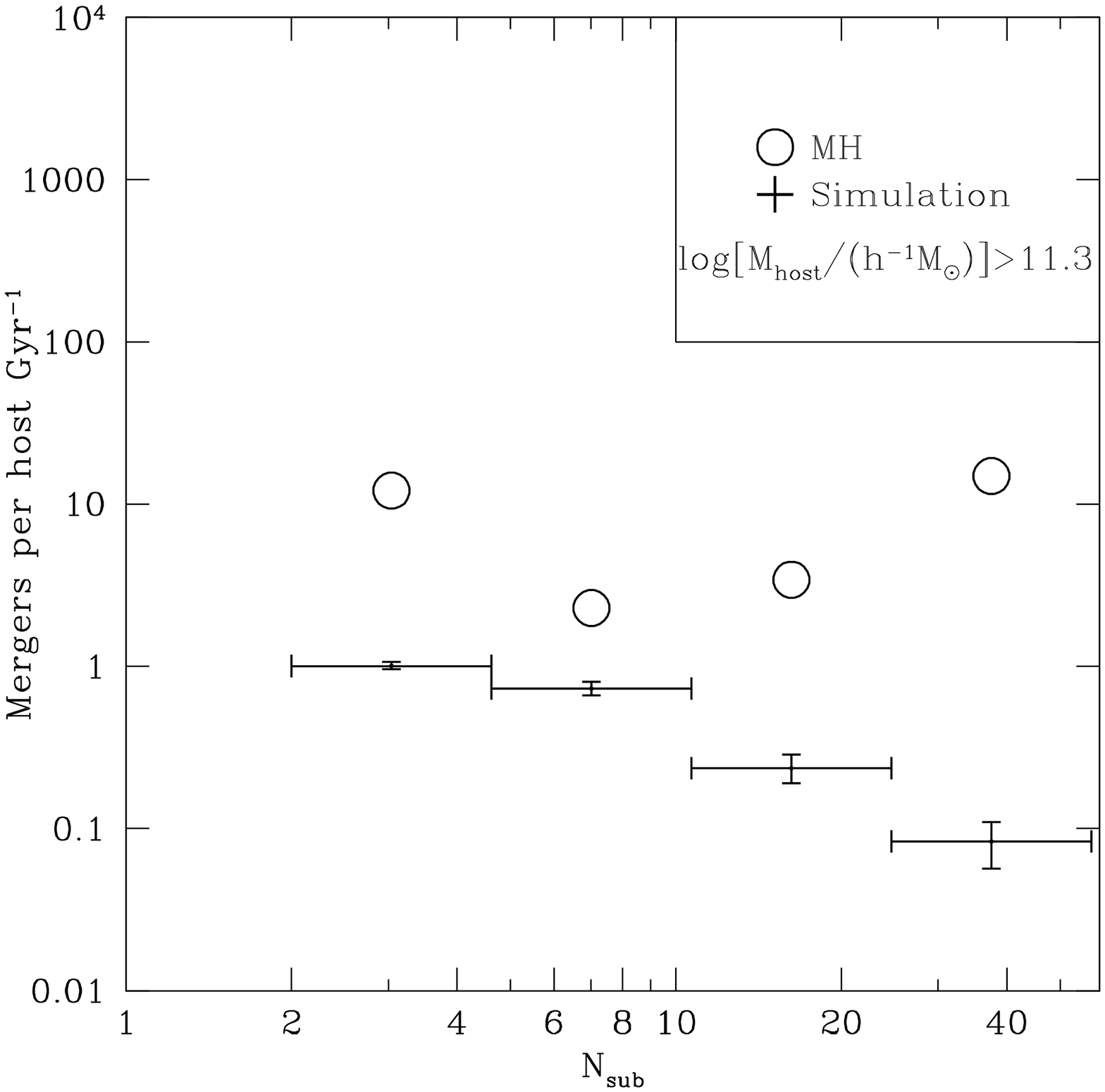}}
{\epsfxsize=2.7 in \epsfbox{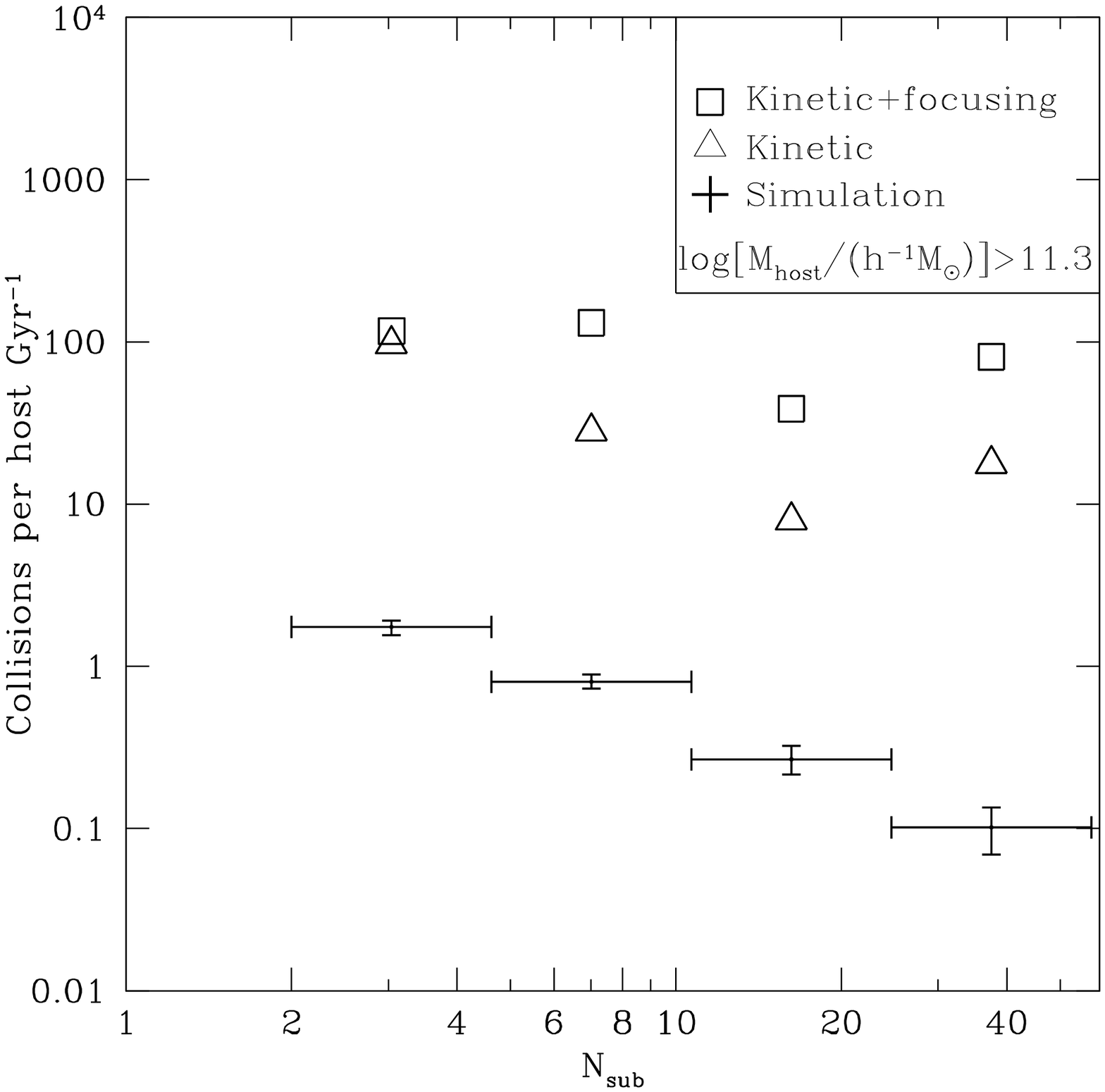}}
\caption{The average number of (a) mergers and (b) collisions per host halo
as a function of $N_{\rm sub}$ (the number of subhalos of the host) for
host halos $>2\times 10^{11} \hsolmass$,
in comparison to analytic models. Symbols are as in Fig. \ref{fig:stat_t}
}
\label{fig:stat_sub_a4}
\end{figure}

\begin{figure}
{\epsfxsize=2.7 in \epsfbox{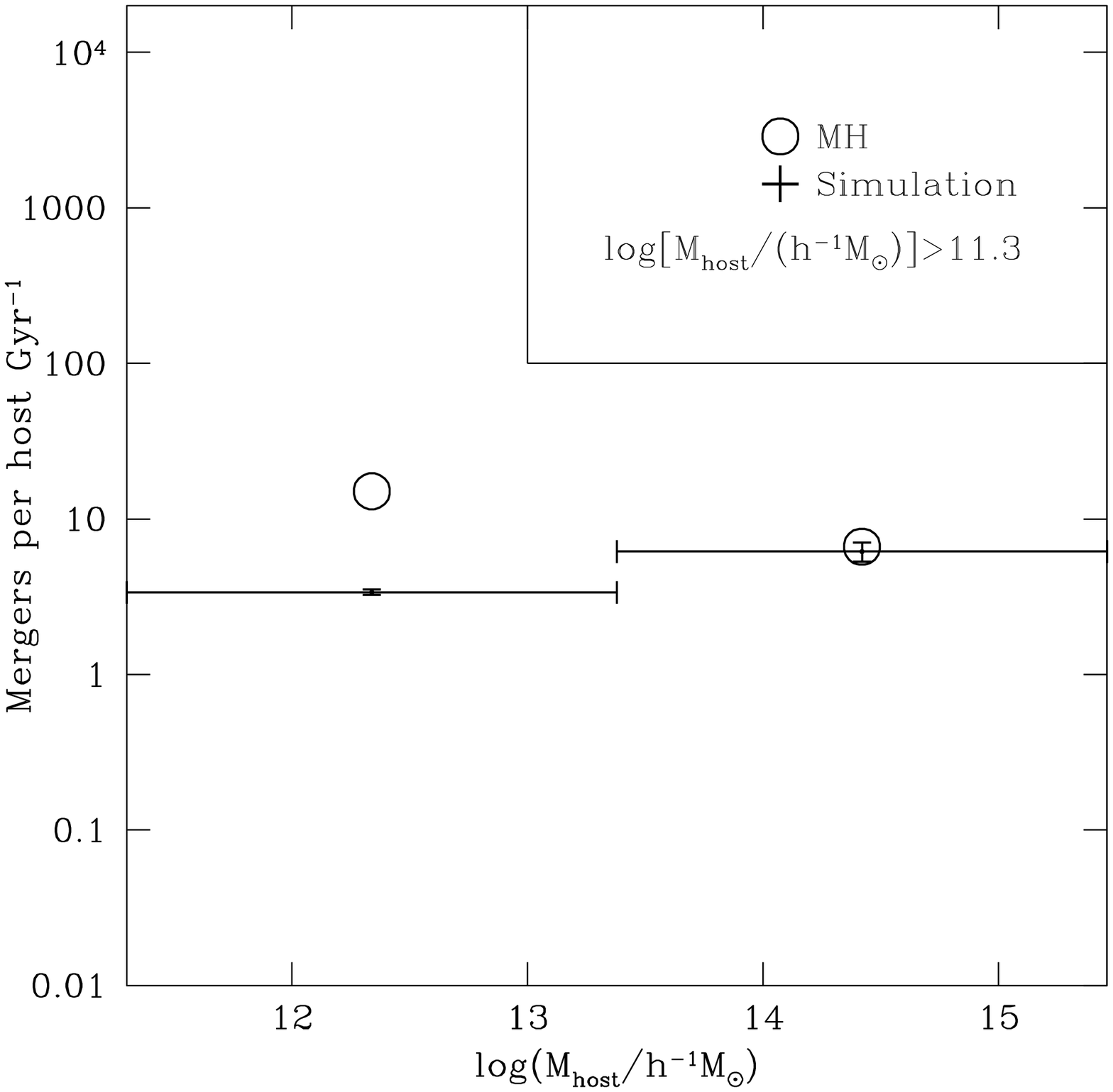}}
{\epsfxsize=2.7 in \epsfbox{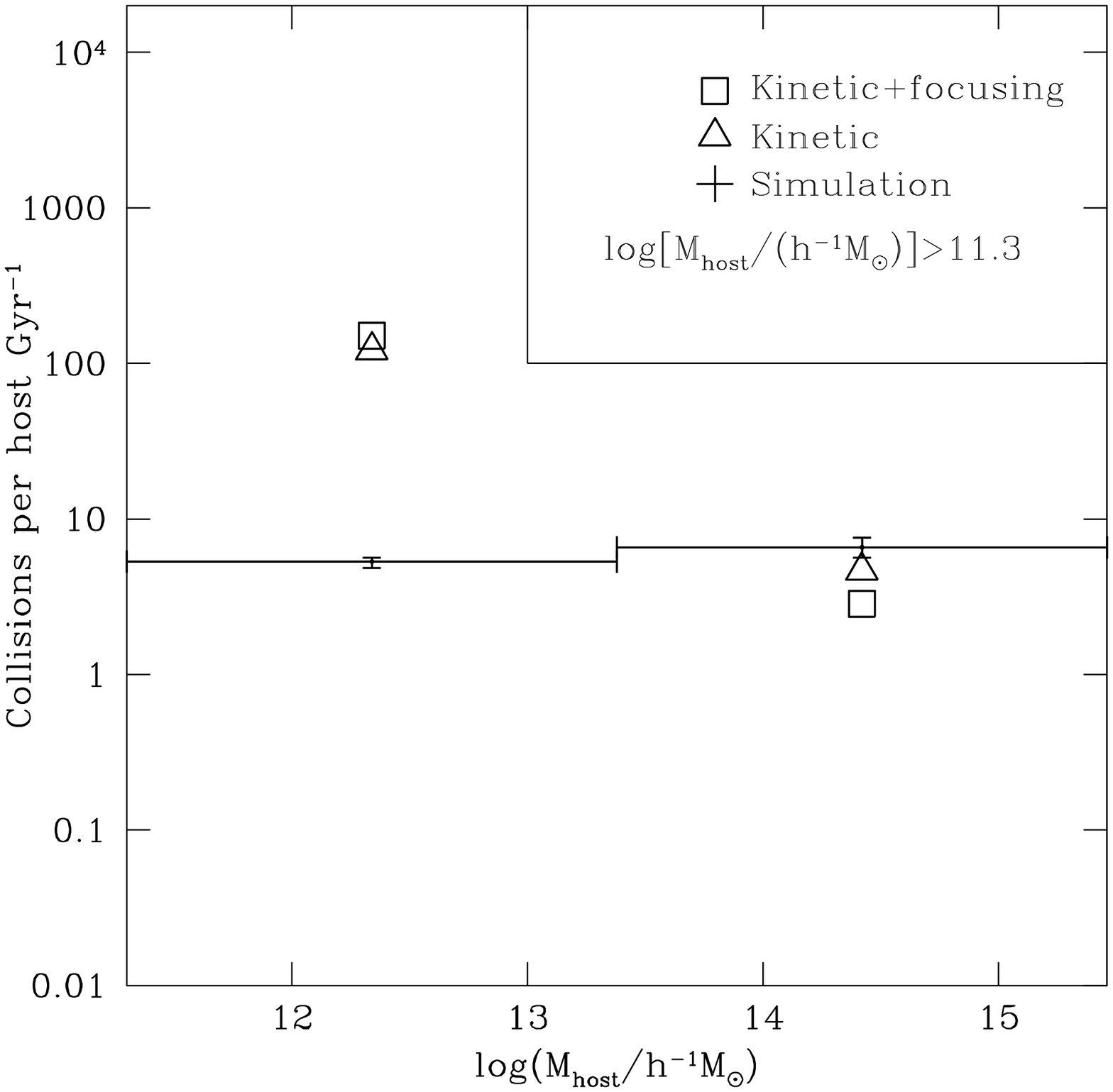}}
\caption{The average number of (a) mergers and (b) collisions per host halo
as a function of the host mass in comparison to analytic
models. Symbols are as in Fig. \ref{fig:stat_t}. The lower
``gravitational focusing" point on the higher mass bin reflects the
different weighting scheme. 
}
\label{fig:stat_mass}
\end{figure}
In an attempt to bin the mass more finely a useful approximation for all
masses of host halos is found to be 
\begin{equation}
{\cal R}_{\rm model} \simeq f(M) {\cal R}_{\rm simulation} \,,
\label{eq:nsv_factor}
\end{equation}
with $\log(f) = -0.27\,\log(M/10^{18.5}\hsolmass)$.
The approximation (\ref{eq:nsv_factor}) works fine for both mass resolutions
when applied to the kinetic theory ($n \sigma \overline{V}$, with or without
gravitational focusing). There is no similar correction formula for the MH
predictions.

In spite of the poorer statistics, we now go back to the evolution of the
collision rate in time (Fig. \ref{fig:stat_t}) and apply the same 
calculation under
the constraint of $M_{\rm host}> 6.3 \times 10^{11}\hsolmass$. This
allows statistically significant results at almost all redshifts. 
Figure \ref{fig:stat_t_11.8} shows the results.
A very nice agreement is obtained for all redshift bins between the MH 
approximation and the simulation. The kinetic theory still overestimates
the collision rate by about an order of magnitude beyond redshift
$z\simeq1$, but gets closer to the simulation results at lower redshifts
if gravitational focusing is ignored.

The MH prediction is 
expected to give somewhat higher
results than the simulation since MH calculations consider the half mass
radius while in the simulation the requirement is of a cross section $\rs$
in conjunction with a final bound state. Figs. \ref{fig:stat_mass}a and
\ref{fig:stat_t_11.8}a show that on average for large host halos
the two merger definitions are similar, but in details and for small halos
they differ. It is encouraging to verify that the merger definition is 
quite robust and
the outcome rate does not depend on its fine details.
The kinetic theory predictions tend to overestimate the rate since
they neglect stripping (and thus shrinking cross section) and harassment
to the degree of disappearance of halos.

\begin{figure*}
\vskip-5truecm
\centerline{\psfig{file=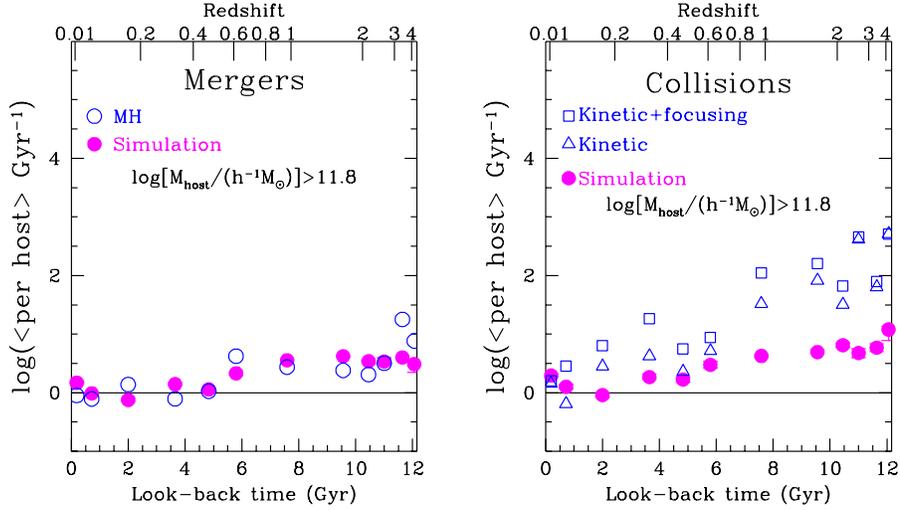
 ,height=13truecm,width=13truecm }}
\caption{
Same as figure \ref{fig:stat_t}, except that here only host halos of mass
$>6.8\times10^{11}\hsolmass$ are included in the comparison to obtain a
much better match with the Makino \& Hut approximation.
Note that for the kinetic theory the discrepancy becomes smaller too.
}
\label{fig:stat_t_11.8}
\end{figure*}

\section{Dynamical friction}
\label{sec:df}

The breakdown of the kinetic theory statistics in the limit
of a few-body system, and
in particular for low masses as seen in 
the simulations, leads us to seek an alternative assessment for the
collision rate in these systems.
Fig. \ref{fig:r_from_par} shows the distance of the identified
collision, $R_{\rm collision}$ from the host center, for all collisions
that involve at least one subhalo. 
Note that beyond the first radial bin, the rise in
the number of collisions as a function of
the radius in which they occur
is slower than expected for a homogeneous distribution throughout the
host's virial volume, as assumed in the kinetic theory approximations
(Eq. \ref{eq:mh}, \ref{eq:rate_coll}).
Instead, the distribution follows more closely a mass profile of $M \propto r$.
In the limit of a binary system, every collision is by definition a host
-- subhalo collision. These are the host -- subhalo collisions that 
contribute to the big spike
in Fig. \ref{fig:r_from_par} near $R_{\rm collision} /\rvir=0$.

Two competing processes are responsible for this type of
collisions: a radial orbit may lead the subhalo directly to the
host's center \cite{moore:98}, 
while a more tangential orbit may end up spiraling
into the center of the host due to dynamical friction.
The characteristic time for spiraling in, as estimated by
dynamical-friction
theory, is clearly relevant for the latter but it may or may not be a
useful
approximation for the former. We therefore set to test the applicability
of the dynamical-friction time scale in the realistic systems of our
simulation.

Traditionally the rate of collisions is estimated using
a simplified form of dynamical friction \cite{kwg:93,sp:99}
which utilizes a very simple approach.
The halo is assumed to be an isothermal sphere, and the subhalo is
assumed to be a point mass on circular orbits, ignoring mass-loss as
it spirals in.
The more elaborate approximations
by \citeN{benson:00} and by Klypin et al. (1999)
allow a more realistic density profile for the halos but they
still assume circular orbits for subhalos.

\begin{figure}
\centerline{\psfig{file=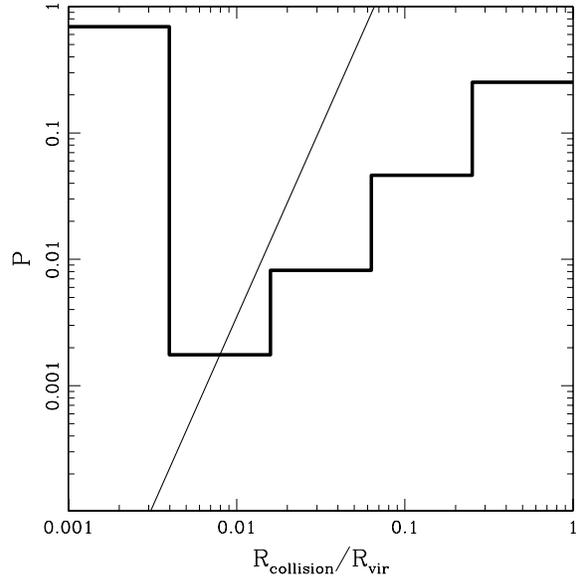
,height=8truecm,width=8truecm }}
\caption{
Distribution of the collision  locations in units of the host $\rvir$.
The diagonal line describes the relative volume in the radial bins,
normalized by the counts at the second inner-most bin.
}
\label{fig:r_from_par}
\end{figure}

We here focus on collisions between subhalos and their hosts. 
The progenitor subhalo of these
collisions is identified in the regime
between $\rvir$ and $\rs$ of the host (apart from the ``cD" case). 
We first check how well 
the traditional dynamical friction timescale
predicts the behavior in the simulation.
The expression for the dynamical friction time in the framework of
NFW density profile for the host halo is given by \cite{BT}
\begin{equation}
\tau^{\rm NFW}_{\rm df} = \int_{r_f}^{r_i} { t_{\rm df}(r) \over
2r} \left({\partial \ln M_{\rm host}(r) \over \partial \ln r} + 1 \right)
{\rm d}r \,,
\label{eq:t_df_nfw}
\end{equation}
where \cite{klypin:overcoming}
\begin{eqnarray}
t_{\rm df} &=& { v_c^3 \over 4 \pi G^2 (\ln \Lambda) M_{\rm sub} \rho(r)
\left[ {\rm erf}(X) - 2Xe^{-X^2}/ \sqrt{\pi} \right] }, \nonumber \\
v_c^2(r) &=& GM_{\rm host}(r)/r;\, \, X=v_c/(\sqrt{2} \sigma_r);\,\nonumber \\
\Lambda &=& {\rvir M_{\rm host}(r) \over r M_{\rm sub}}\,,
\end{eqnarray}
and $\sigma_r$ is the one-dimension velocity dispersion at radius r,
\begin{eqnarray}
\sigma^2_r &=& v_{\rm max}^2{2x(1+x)^2\over
f(2)}\int^{\infty}_x{f(x)\over x^3(1+x)^2}dx, \nonumber \\
f(x) &=& \ln(1+x) - {x \over 1+x}\,\,; x = {r \over R_{\rm s}} \,\nonumber \\
v_{\rm max} &=& v_c(2.15R_{\rm s}) \,.
\end{eqnarray}

The halo profile, and therefore the circular velocity $v_c(r)$, are
assumed to remain constant between subsequent time-steps.
The radii $r_i$ and $r_f$ are the initial and final radii to consider,  
respectively.
The initial separation $r_i$ is identified at the earlier redshift as
the offset between the center of mass of the host and that of the subhalo. 
The final radius $r_f$ is the sum of the $\rs$ values for the host and
the subhalo, to maintain consistency with the collision definition.
For the ``cD" case $r_f$ is taken to be zero.
The choice for the subhalo mass is somewhat ambiguous: the scheme we use
{\it modeles} the subhalo mass according to its NFW density profile. An
alternative is to consider only the mass within $R_{\rm t}$ (c.f.
\S\ref{subsec:model}). We tried both possibilities and found only
negligible differences between the two results. Note, however, that the subhalo
mass does change in between the two time-steps and thus deviations from the
no-mass-loss approximation (Eq. \ref{eq:t_df_nfw}) are expected. 
Interestingly, since the subhalo mass usually
decreases, the approximation should underestimate the dynamical friction time. 
We shall soon see that the opposite happens, in many cases, indicating
that the simplified model may be crude for other reasons too.

An alternative approach would be to calculate $\tau_{\rm df}$ in a
similar way it is done in certain semi-analytic models (e.g., Somerville \&
Primack 1999). 
The host halos are
assumed to follow a singular isothermal sphere (SIS) profile whose
constant circular velocity is given by the virial mass ad radius:
$v_c = \sqrt{ G M_{\rm vir}/R_{\rm vir}}$. The dynamical friction time
scale is then simplified to \cite{BT}

\begin{equation}
\tau^{\rm SIS}_{\rm df}  \approx {1.17 \, (r_i^2 - r_f^2)\, v_c \over
 \ln  \Lambda \, G \, M_{\rm sub}} \, ,
\label{eq:t_df_sis}
\end{equation}
where the Coulomb logarithm is approximated by
$\Lambda =M_{\rm host} / M_{\rm sub}$.

Among all collisions that have taken place between $z_2$ and $z_1$ we
select only collisions of the type subhalo -- host, namely systems 
that at $z_2$ already belonged to the same virialized system
where approximations for dynamical friction time are applicable.
We identify the progenitor that at $z_2$ used to be the
subhalo and calculate its expected $\tau_{\rm df}$ according to Eqs.
(\ref{eq:t_df_nfw}) and (\ref{eq:t_df_sis}). 
We do not take its actual orbital parameters into account,
because we are interested in finding out whether $\tau_{\rm df}$
is a good approximation when applied to a statistical sample of
collisions as in the semi-analytic models.
We then compare the expected dynamical friction time of this
individual collision with the elapsed time $\Delta t$ between $z_2$ and
$z_1$ in the simulation.
The time resolution of the analyzed simulation
output times does not allow a
detailed comparison of the calculated dynamical friction time and the
collisions as they happened in the simulations. The proper time gap between
successive stored times serves instead as an upper limit for the real
time.

If the dynamical friction time is a reliable estimate, all
the individual $\tau_{\rm df}$ values should be smaller than $\Delta t$,
because we selected for all collisions that have actually happened
on a time shorter than or equal to $\Delta t$.
The choice of going ``backwards" in time, namely from identified
collisions to their progenitors, is made because some of the subhalos at
the higher redshift $z_2$ got disrupted by the time of the
lower redshift $z_1$ and can no
longer serve as dynamical friction probes.

\begin{figure}
 {\epsfxsize=3.7in \epsfbox{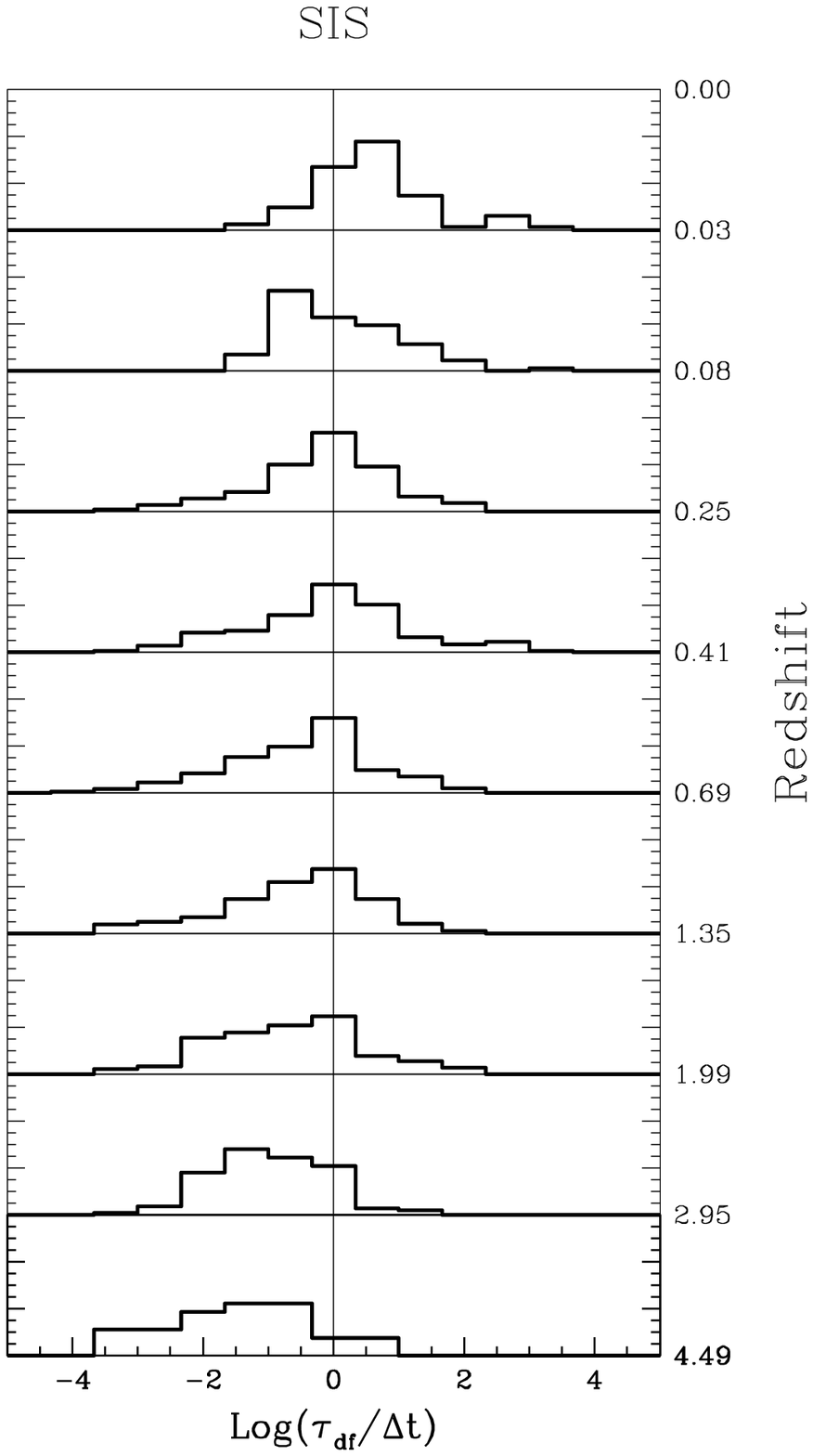}}
 {\epsfxsize=3.7in \epsfbox{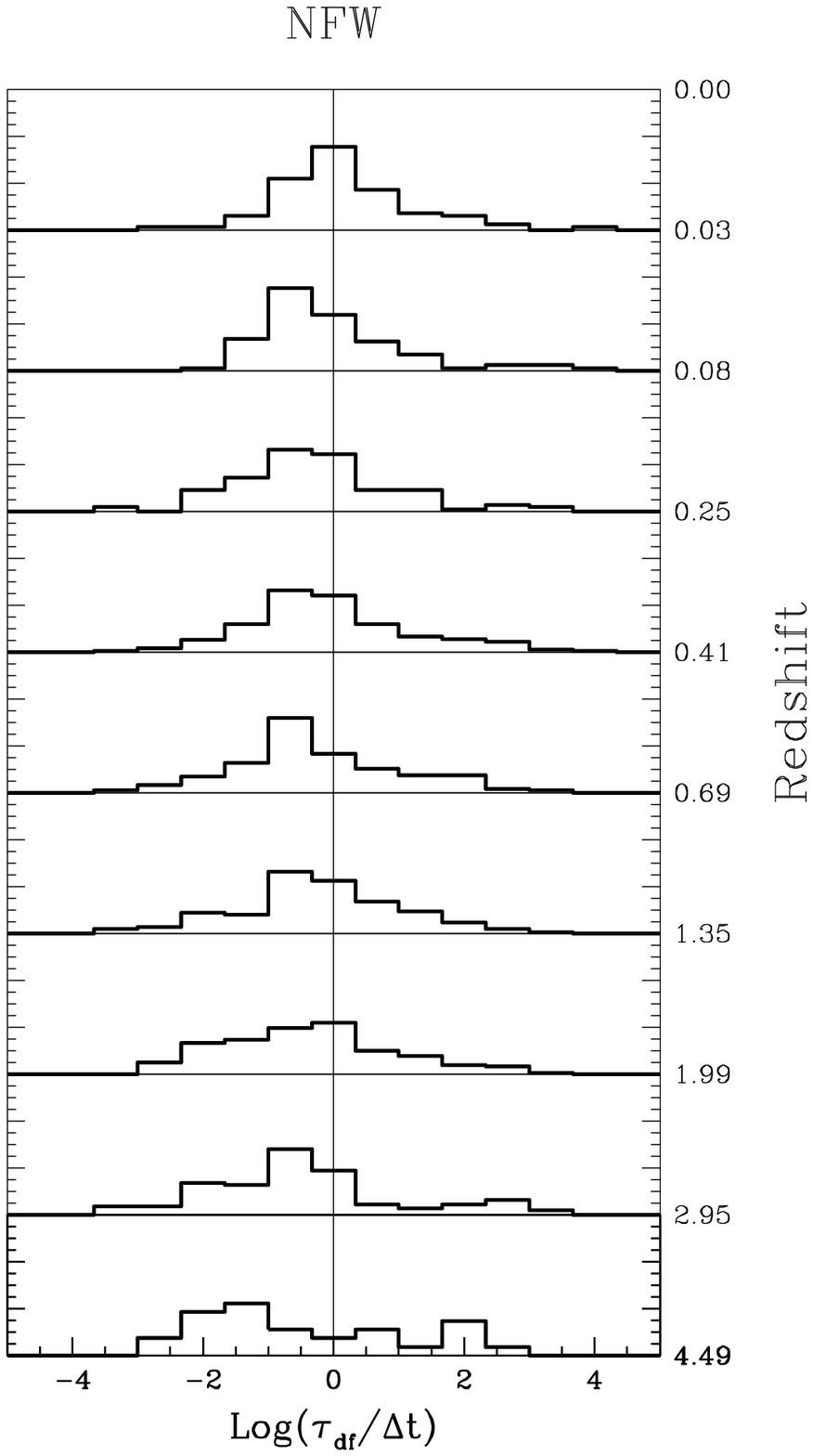}}
\vskip-0.5truecm
\caption{Distribution of dynamical friction timescale $\tau_{\rm df}$
in units of the elapsed time $\Delta t$ between subsequent simulation
outputs for various redshifts (right).  Here $\tau_{\rm df}$ is
calculated assuming that the host halo radial profile is either singular
isothermal sphere (top panel) or NFW (bottom panel). }
\label{fig:t_df}
\end{figure}

Figure \ref{fig:t_df} shows the distribution of dynamical friction time
in units of $\Delta t$,
for all progenitor subhalos in the simulation at a given redshift, 
and for 9 timestep pairs.
For the SIS calculation
the dynamical friction time gradually becomes more of an overestimate as
the simulation progresses in time.
At high redshift $z\ga 1$, more than $65\%$ of all subhalos spiral
into their hosts in a time comparable to or longer than their calculated
SIS dynamical friction time. 
At low redshifts, the SIS dynamical friction is typically an overestimate, 
sometimes by more than two orders of magnitude.
The use of the best-fit NFW profile 
brings about a somewhat worse agreement at high redshift. This is due to
the lower halo concentration at high redshift that causes subhalos to 
be classified as ``cD",  residing within the $\rs$ of their hosts.
The definition of $r(t_f)$ in such cases is more ambiguous and may not
reflect the exact collision radius. 
Moreover, at high redshift a ``cD" case
may occur for progenitors of rather similar masses 
where the dynamical friction approximation breaks down anyway.
In the SIS calculation for the ``cD" cases, the steeper density profile 
near the center ($\propto r^{-2}$) tends to shorten the dynamical 
friction time estimate and partly compensates for the method inaccuracies.
However at low redshift, where such degeneracies are rare,
the NFW estimate does better than the SIS
estimate and by $z\simeq0$ it overestimates the time for only $\sim50\%$ of all
cases, in comparison to the $\sim70\%$ in the SIS case. 

This spread of $\tau_{\rm df}/\Delta t$ should partly be attributed 
to the fit errors, as at
high redshift the halos tend to be less concentrated 
(BKS+)
and the calculated dynamical friction time is
affected at large radii by the modeled $\rs$. The bigger errors
for $\rs$ at high redshift propagate thus to an error in the
dynamical friction time estimate.

Semi-analytic models that use dynamical time estimates to
predict mergers must take the errors in these estimates
into account or else they
underestimate the merger and collision rate both in absolute terms and more
severely at low redshift relative to high redshift.

Figure \ref{fig:t_df_longer} summarizes Fig. \ref{fig:t_df} and
depicts the fraction of mergers between subhalos and hosts that 
occurred faster than the
dynamical friction time prediction as a function of redshift.
\citeN{kk:99} have analyzed the dynamical friction time-scale
for {\it clusters} in these simulations using a different methodology,
and found that at high redshift the
dynamical friction time-scale truly represents the cluster
evolution whereas at lower redshift for a substantial fraction of the
subhalos the dynamical friction timescale is longer than the actual
merging timescale. 

\begin{figure}
\centerline{\psfig{file=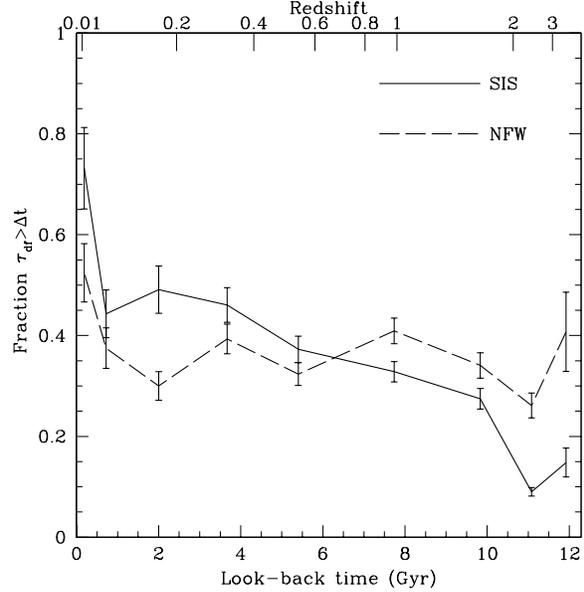
 ,height=8truecm,width=8truecm }}
\caption{Fraction of collisions where dynamical friction timescale
overestimates the collision time as a function of look back time.  }
\label{fig:t_df_longer}
\end{figure}

Some of the mismatch between the $\tau _{\rm df} $ and $\Delta t$
may arise due to the invalidity of the circular orbit assumption. 
If most collision progenitors had initially radial orbits  then
the collision time should be smaller. 
The redshift dependence of the discrepancy can then be explained in
terms of bigger infall velocities at lower redshift and domination of
radial orbits near big groups and clusters.

\begin{figure}
{\epsfxsize=2.7 in \epsfbox{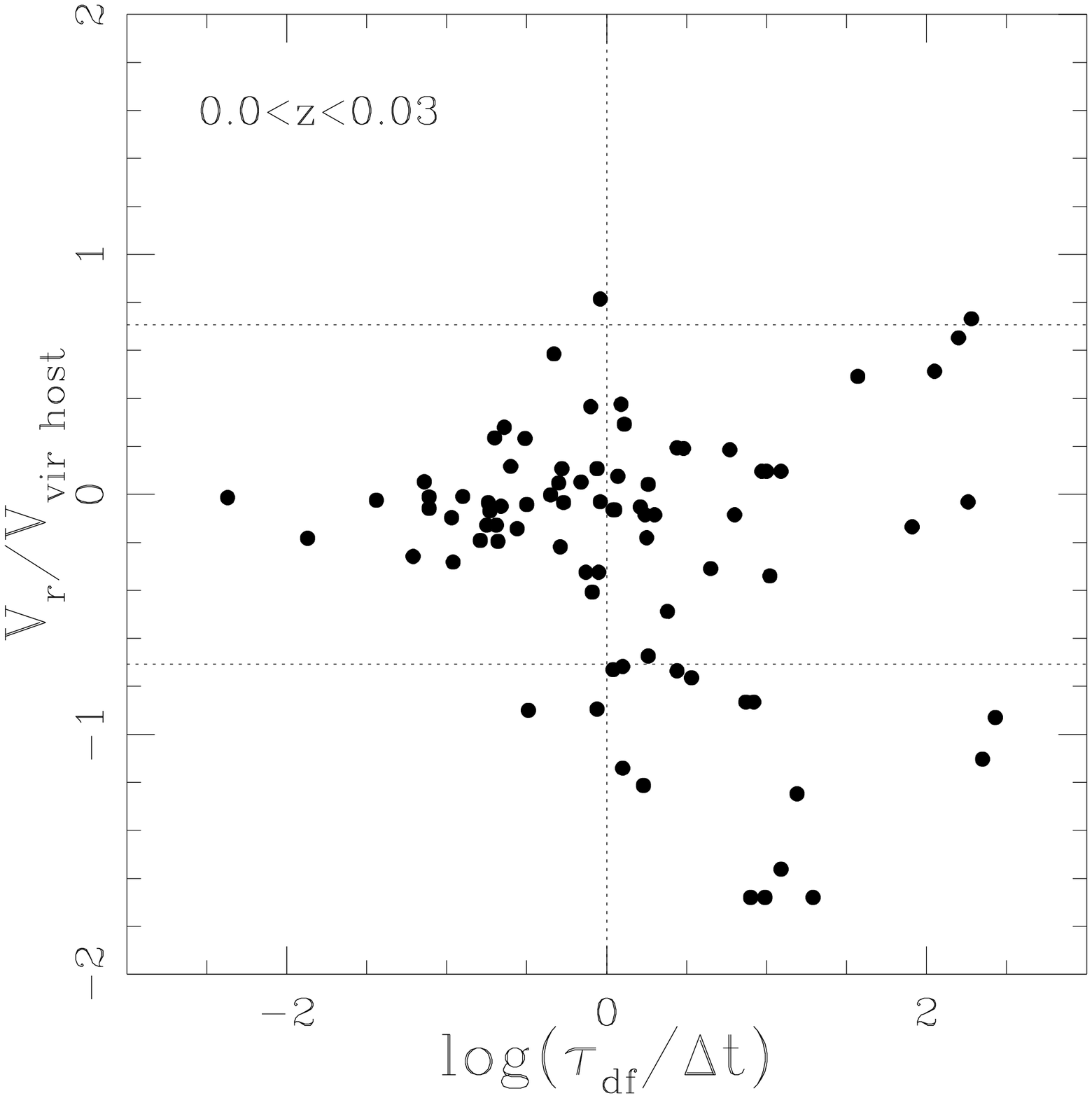}}
{\epsfxsize=2.7 in \epsfbox{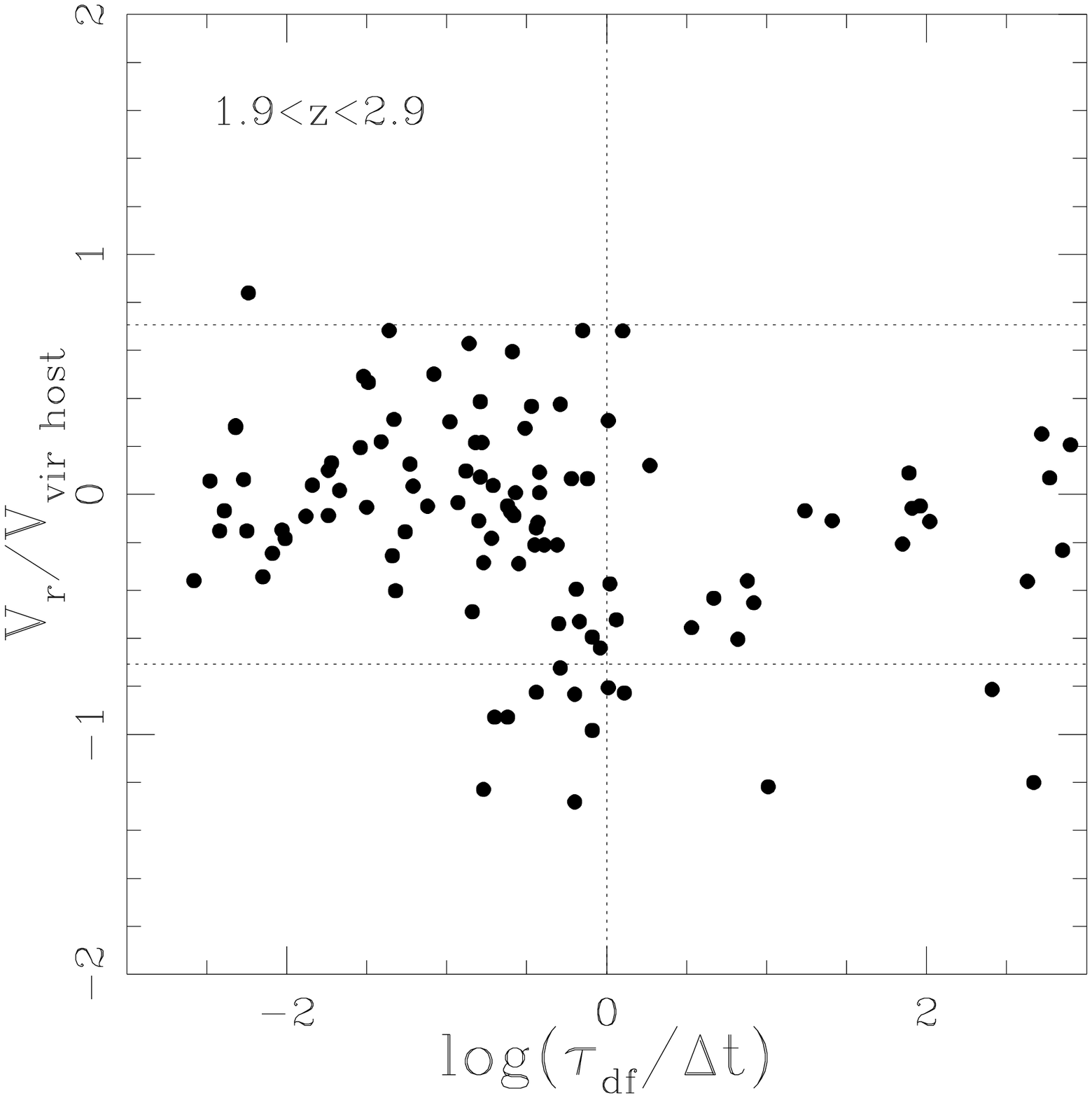}}
\caption{ Radial velocity in units of the host virial velocity 
as a function of the deviation from NFW dynamical friction time estimate. 
The dotted lines mark the $2^{-1/2}$ ratio.
For collisions between $0.03>z>0$
(top) and $2.9>z>2.3$ (bottom).} 
\label{fig:t_df_deviation}
\end{figure}

Figure \ref{fig:t_df_deviation} (top) shows the radial velocity of the subhalo
progenitor normalized to the modeled circular SIS velocity (or
equivalently $v_{\rm vir}$ of NFW) 
at the lower redshift. The radial velocity should
correlate with deviations from the dynamical friction time predictions,
especially the negative radial velocities. 
Indeed,  about $40\%$ of the mergers for which the dynamical friction
estimate failed to yield the right time scale, show radial velocities
$\vert v_{\rm r} \vert > v_{\rm vir}/ \sqrt{2}$, 
whereas this fraction is
only $\sim 7\%$ for mergers of correct time scale prediction.
Interestingly, there is a marginal indication
that even positive radial velocities correlate with the breakup of the
approximation. 
This may be interpreted as orbit eccentricity being an important
reason for the deviation. 
At higher redshift (Fig. \ref{fig:t_df_deviation}, bottom)
progenitors have more circular orbits (only $12\%$ and $20\%$ of
$\vert v_{\rm r}\vert  > v_{\rm vir}/ \sqrt{2}$ 
for correct and deviating mergers respectively) 
and therefore the deviation decreases.
The difference between the time estimates by the SIS and NFW models
is gradually amplified at low redshifts due to the larger concentration
there (BKS+); the deviation of the simplified SIS profile from
the realistic NFW profile grows with $c_{\rm vir}$, which makes the SIS a weak
approximation for high-$c_{\rm vir}$ halos.

Figure \ref{fig:t_df_r1} indicates that the deviations from the dynamical
friction time estimate
are correlated with the location of the progenitor
subhalo (as identified in the output time just preceding the merger).
Subhalos from the outskirts of the host tend to merge
on a timescale much shorter than the dynamical friction time estimate.
The discrepancies stem from such progenitors because they either 
spend longer radial oscillation time there, or because they have just entered 
the host halo and are not yet virialized. 
A counter effect comes
from deflections by other subhalos in the same host halo, which tend to
make the dynamical friction timescale an underestimate of the true
spiraling-in time \cite{merritt:83}.
Apparently this effect is overruled by the effects working in the
opposite direction.

\begin{figure}
\centerline{\psfig{file=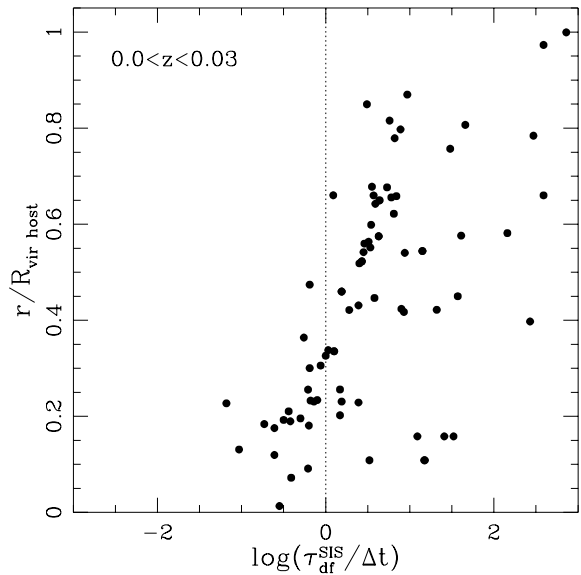
 ,height=8truecm,width=8truecm }}
\caption{
Location of the subhalo progenitor (in units of the host $\rvir$) as
function of the deviation from dynamical friction time estimate (SIS).
}
\label{fig:t_df_r1}
\end{figure}

Another possible cause for the mismatch between estimated and observed
$\tau_{\rm df}$ may be the breakdown of the $M_{\rm sub}/M_{\rm host} \ll 1$
assumption used in the derivation of $\tau_{\rm df}$. We checked
for a correlation between
the deviation from dynamical friction time
estimate and the subhalo -- host mass ratio. 
For the SIS estimate there does not seem to be any correlation 
between these two quantities. However all the deviating systems host
-- subhalo at high redshift are systems for which $M_{\rm sub}/M_{\rm
host} > 0.1$ (the opposite is not true). Again, at high redshift when
the concentration is low, and the subhalo mass is distributed over
large radius, the assumption about a point-like subhalo with a fixed mass
that does not affect the background potential breaks down, and
hence the dynamical friction formula ceases to be valid especially for similar
masses of host halos and the spiraling subhalo.
This effect is somewhat stronger when the actual density profile (i.e., NFW)
is taken into account.
Despite the fact that in the dynamical friction framework these effects
tend to weaken the dynamical friction and thus make the dynamical
friction time estimate longer, other processes, e.g., two body
relaxation are more common for ``puffed" halos and may work to shorten the
spiraling-in time scale.
Note that by moving from the modeled $M_{\rm sub}$ to $M_{\rm
sub}(R_t)$, $\tau_{\rm df}$ becomes longer ($M_{\rm sub} \ga
M_{\rm sub}(R_t)$) and thus the choice of using $M_{\rm sub}$
cannot be the reason for the deviation.

\section{Summary}
\label{sec:conc}

We investigated the interaction rate of subhalos, which
should be an important ingredient in the process of galaxy formation.
This study is needed because virialized systems of objects evolve differently 
than unvirialized systems in which the objects are only weakly correlated.
Also, the observed large fraction of galaxies in groups suggests that
virialized systems of two galaxies or more are very common,
and such high-density environments tend to be preferentially selected 
for observation.
We thus computed the collision and merger rates within 
virialized systems in N-body simulations, and used the results to evaluate
useful semi-analytic approximations.
The large sample of halos of a range of masses and multiplicities
enabled a comprehensive statistical analysis, 
and the especially designed halo finder and collision identifier allowed 
a proper quantitative study of substructure evolution.
Given the resolution limits of our simulations and halo finder,
we could account for only the two 
most massive members of a group like the Local Group,
but since these two members commonly dominate the group 
dynamics, the mass resolution is adequate for our purpose.

We found that the comoving collision rate of subhalos is a substantial fraction
to the total collision rate of halos in the $\Lambda$CDM simulation.
At $z\sim 4$ the fraction of collisions that involve substructure
is $\sim 0.3$, and this fraction grows 
to $\sim 0.75$ at low redshifts (Fig. \ref{fig:rate_sub} (top)).  In physical
coordinates, this corresponds to a collision rate $\propto (1+z)^{3-4}$
(Fig. \ref{fig:rate_sub_phys}).
The evolution of the subhalo collision rate follows that of the total 
collision rate, namely a
sharp rise between $z=4$ and 2, followed by a decline between $z=1$ and 0.4. 
(Fig. \ref{fig:rate_sub_sub}). Even though the exact evolution is affected by
the specific cosmology simulated and the results are valid for a
specific mass range limited by the resolution of the simulation,
the general behavior of a rise followed by a decline should be robust.
We find that the collision rate of subhalos lags behind the total rate 
by about $1-2$ Gyr. It also exhibits a different behavior at low redshift, 
when many large virialized systems exist and contribute a significant
fraction of the collision rate and a rise in the merger rate.

The mass function of progenitors that participate in
subhalo mergers is somewhat different from that of distinct halo collisions 
(Fig. \ref{fig:mratio_sub}), 
amplifying small differences in the tails of the mass functions for 
subhalos and distinct halos.
The fraction of ``major" interactions among those which
involve at least one subhalo is only $\la 0.5$, unlike the $\sim 0.6$
fraction of major interactions among distinct halos.
The moral is that accurate 
theoretical predictions of interaction rates should not be  
solely based on the halo mass function as derived from approximations
such as Press-Schechter or Extended-PS, which ignore substructure. 

We found that the Makino-Hut formula can serve as a reasonably good
approximation for subhalo merger rates, as long as host halos more massive
than $\sim 6 \times 10^{11}\hsolmass$ are considered. At lower masses, the MH
formula overestimates the merger rate by
an order of magnitude. At high redshift, where less massive halos are
abundant, the MH formula ceases to be a good description for the 
merger rate of the subhalo population.

Predictions of the subhalo collision rate from kinetic theory are found to
recover the actual rate for host halos more massive than $\sim
10^{13.5} \hmsun$, but fail to do so, by one or two orders of magnitude,  
for less massive host halos, depending on their multiplicity. 
We presented a correction formula to match kinetic theory predictions with the
simulated halos, using either the mass of the host, or the number of subhalos
it contains. 

For interactions between subhalos and their hosts,
we examined the accuracy of the dynamical friction timescale as estimated
for circular orbits.
In general, the estimate based on the NFW density profile is acceptable
for about half the cases.
It significantly overestimates the collision time in about
$40\%$ of the cases,
and by more than an order of magnitude in $\sim 20\%$ of the cases.
One reason for this failure is that, at least 
at early times, the halos tend to be of low concentration
and the high-mass host halos are of low abundance,
making the assumption of $M_{\rm sub}/M_{\rm host} \ll 1$
invalid in many cases.
The deviation from circular orbits is clearly another reason for the
failure
of the dynamical-friction approximation, as demonstrated at the most
recent output time of the simulation.
We find that, by some coincidence, the estimate based on the approximation 
of halos as isothermal spheres is acceptable at high redshift.
This is good news for semi-analytic models that commonly use this
approximation. 
At low redshifts, on the other hand,
the concentration tends to be high, and the SIS predictions become poor.
The moral is that, in order to mimic substructure dynamics in a cosmological
context, one needs to appeal to accurate simulations, or use approximations 
that were derived from such simulations.
We have reasons to assume that these approximations, once expressed 
properly in terms of the host-halo mass, can be robust
and not too sensitive to the particular cosmology.
Proper collision rates can then be incorporated into galaxy formation
scenarios and connect them to the evolution of large-scale structure.

\section*{Acknowledgments}
The simulations were performed at NRL and NCSA.  This work was
supported by grants from NASA and NSF at UCSC and NMSU, and by 
Israel Science Foundation and US-Israel Binational Science Foundation
grants.  JRP gratefully acknowledges a Forchheimer Visiting
Professorship at The Hebrew University.

\bibliographystyle{mnras}
\bibliography{mnrasmnemonic,all_refs}
\end{document}